\documentclass[english]{article}
\usepackage[utf8x]{inputenc}
\usepackage[english]{babel}
\usepackage{amsmath,amssymb,amsfonts,amsthm,enumitem,array,mathtools}
\usepackage[math]{iwona}
\usepackage[T1]{fontenc}
\usepackage{hyperref}
\usepackage{comment}
\usepackage{tikz-cd}
\usepackage{stmaryrd}
\usepackage{faktor} %per gli insiemi quoziente
\usepackage{bbm}
\usepackage{mathtools}  % for paired delimiters 
\usepackage{mathrsfs}
\usepackage{cancel}
\usepackage{indentfirst}
\usepackage{pgfplots}
\usepackage{tikz}
\usetikzlibrary{shapes}
\usetikzlibrary{plotmarks}
\usepackage{cite}
\usepackage{cleveref}
\Crefname{subsection}{Subsection}{Subsections}
\usepackage[numbers]{natbib}
\usepackage[margin=2.5cm]{geometry}

\usepackage[Lenny]{fncychap}        
    \ChTitleVar{\Huge\bfseries}
\usepackage{titlesec}

\newtheorem{lem}{Lemma}[section]

\newtheorem{prop}{Proposition}[section]

\newtheorem{thm}{Theorem}[section]
\newtheorem*{thm*}{Theorem}

\newtheorem{rem}{Remark}[section]

\newtheorem{ex}{Example}[section]

\newtheorem{notat}{Notation}[section]

\newtheorem{defn}{Definition}[section]

\newtheorem{constr}{Construction}[section]

\makeatletter
\newenvironment{myproof}[1][\proofname]{\par
    \pushQED{\qed}%
    \normalfont \topsep6\p@\@plus6\p@\relax
    \trivlist
    \item[\hskip\labelsep
        \scshape
        #1\@addpunct{.} ]\mbox{}\nobreak}
    {\popQED\endtrivlist\@endpefalse}
\makeatother
\usepackage{lipsum}

\newcommand{\Div}{\mathrm{Div}}
\newcommand{\Supp}{\mathrm{Supp}}

\newcommand{\Spec}{\mathrm{Spec }}

\newcommand{\Pic}{\mathrm{Pic}}

\newcommand{\DDiv}{\mathrm{Div}}

\newcommand{\OO}{\mathcal{O}}

\newcommand{\Sp}{\mathrm{sp}}

\newcommand{\ev}{\mathrm{ev}}
\newcommand{\ddiv}{\mathrm{div}}

\newcommand{\N}{\mathbb{N}}

\newcommand{\Z}{\mathbb{Z}}
 
\newcommand{\C}{\mathcal{C}}

\newcommand{\E}{\mathcal{E}}
\newcommand{\F}{\mathbb{F}}
\newcommand{\p}{\mathbb{P}}

\newcommand{\ie}{\textit{i.e.,~}}
\newcommand{\eg}{\textit{e.g.,~}}

\setenumerate{label*=\arabic*.}

\makeatletter
\newcommand{\myitem}[1]{%
\item[#1]\protected@edef\@currentlabel{#1}%
}
\makeatother

\newcommand{\funcDef}[5]{
\begin{array}{cccc}
{#1:} & {#2} & \longrightarrow & {#3}\\
     & {#4} & \longmapsto & {#5}
\end{array}}

\title{Locally recoverable codes from elliptic surfaces with availability and hierarchical locality}
\author{Elena Berardini \thanks{CNRS; IMB, University of Bordeaux, France. Email: elena.berardini@math.u-bordeaux.fr} \and Andrea Fornetto \thanks{Email: a.fornetto@studenti.unipi.it}}
\date{}

\begin{document}
\maketitle
\begin{abstract}
    In this paper, we propose several constructions of Locally Recoverable Codes from elliptic surfaces. In particular, we are able to obtain codes with availability $t>2$, codes with hierarchical locality and, finally, codes which combine availability and hierarchical locality. Our constructions rely on the properties of the torsion groups of elliptic curves and on the fibered structure of elliptic surfaces. In particular, the geometry of the surface is used to introduce a multi-dimensional setting, allowing for more recovery sets, eventually nested one within another.
\end{abstract}
\setcounter{tocdepth}{1}
\tableofcontents
\section{Introduction}
In the last decade, the development of distributed cloud storage systems has motivated the search for methods to improve the resilience of the storage infrastructures against data loss. To achieve this goal, Locally Recoverable Codes (LRCs) were introduced \cite{TamoBarg2014}: a linear code $\C$ is said to have \emph{locality} $r$ if, for every symbol in a codeword, there exists a recovery set of $r$ coordinates such that the symbol can be recovered from the symbols at those coordinates whenever it is erased. 
In the context of cloud storage, this property allows the information on an unavailable server to be retrieved from a small set of other servers.

It is possible to equip LR codes with two main additional properties: \textit{availability} and \textit{hierarchical locality}. An LRC $\C$ has availability $t \geq 2$ if every symbol has $t$ disjoint recovery sets. On the other hand, $\C$ is a hierarchical LRC (HLRC) if every symbol has at least two nested recovery sets and the hierarchical structure consists of the full code, the middle codes and the lower codes. All these features can be achieved, for instance, with well-chosen Algebraic Geometry (AG) codes, exploiting the arithmetic properties and rich geometry of algebraic varieties.  
In the literature, this solution has been primarily addressed by working on algebraic curves (see, \eg \cite{BargTamoEtAl2017,Malmskog2018}). Subsequently, there have been some developments in the construction of LRCs from algebraic surfaces, with a particular focus on fibered surfaces \cite{BargEtAl2017,SVV, BergEtAl2024, SalVic25, BlaHal25}.

The main goal of this paper is to propose a construction of LRCs codes from surfaces which combines availability and hierarchical locality. Elliptic curves and elliptic surfaces play the main role in our setting: the group of points on the former and the fibered structure on the latter furnish us with two relevant tools for obtaining the features we want.

\paragraph{Contributions.} The main contributions of this paper on LR codes from elliptic surfaces are summarized below.
\begin{itemize}
    \item \textbf{High availability on elliptic surfaces via torsion subgroups.} The $m$-torsion subgroups of the elliptic fibers allow us to achieve availability $t=m > 2$ (see \Cref{prop: hier_availability_ell_surfaces}), by interpreting the torsion group as a direct product of cyclic subgroups of order $m$, generalizing the work made in \cite[Section VI]{SVV}. Indeed, we obtain a Cartesian grid on the evaluation set of each elliptic fiber which, applying Abel's Theorem, guarantees the existence of several disjoint recovery sets for each point, avoiding any restriction on the dimension of the code.
    \item \textbf{Hierarchical locality on elliptic surfaces through formal lifting via the generic fiber.} We obtain hierarchical locality on elliptic surfaces by exploiting the same property from codes on elliptic curves (see \Cref{prop: parameters_surfaces section 4}). The transition of the constructions from elliptic curves to elliptic surfaces is performed using the formal definition of the generic fiber and the specialization maps.
    \item \textbf{Combined hierarchical locality and availability on elliptic surfaces.} The simultaneous realization of availability and hierarchical locality on elliptic surfaces (see \Cref{prop:final_params}) is made possible by combining the two approaches introduced above. While the combination of these two properties has been achieved in the literature before by evaluating functions over fibered products of curves (\cite{ballentine2019}), their coexistence on higher dimensional varieties, such as surfaces, was unexplored until now, to our knowledge. 
\end{itemize}

In all our code constructions, we carefully design the spaces of rational functions through towers of quotients and Riemann--Roch spaces, obtaining good control on the function evaluations at the points of the evaluation set. We avoid restrictive constraints on the choice of the evaluation set, while increasing the minimum distance of the resulting codes.

\paragraph{Organization of the paper.} In Section \ref{sec:setting}, we recall background notions used throughout the work. We start with fundamental definitions from coding theory, focusing on LRCs, availability, and hierarchical locality, together with the known estimates for the parameters. The second part of the section contains essential results from algebraic geometry, concerning quotients of curves, elliptic curves and elliptic surfaces, leading to a brief overview of AG codes. In Section \ref{section: availability ell surfaces}, we present our construction of LRCs with availability from elliptic surfaces, generalizing the work done in \cite{SVV}. In Section \ref{section: HLRC elliptic surfaces}, we start building a hierarchical structure on elliptic curves using sequences of quotient isogenies. Then, we discuss the lifting of these codes to elliptic surfaces via the generic fiber and the specialization maps. In Section \ref{section: HLRC with availability}, we present our final construction, obtaining LR codes with hierarchical locality and availability $t=2$ at each level of the hierarchy, mainly revising the construction obtained in the previous section, adding more recovery sets. For each of the aforementioned constructions, we prove bounds for the resulting parameters. Finally, Section \ref{sec: conclusion} contains some concluding remarks, together with the presentation of potential further developments of our work.

\section{Prerequisites}\label{sec:setting}
In this section, we aim to introduce all the notions which will be useful for understanding the motivation and results of the present paper, both on the coding theory side and on the algebraic geometry side. The reader who is already at ease with the definition of locally recoverable codes and their properties, and with the theory of algebraic curves and surfaces, and algebraic geometry codes, can skip this first section.
\smallskip

For the rest of the paper,  we let $p$ be a prime number and $q=p^h$, and we consider the finite field $\F_q$.

\subsection{Coding theory}

   A linear \textit{code} is a linear subspace $\mathcal{C} \subseteq \F_q^n$ for some $n \in \N$. Every string contained in $\C$ is called \textit{codeword}, whose components are named \textit{symbols}. Moreover, the parameters \textit{$n$} and \textit{$k= \dim_{\F_q}(\C)$} are usually respectively called \textit{length} and \textit{dimension} of the code $\C$. 
   Given two vectors $v=(v_1, \ldots, v_n)$ and $w=(w_1, \ldots, w_n)$ in $\F_q^n$, their \textit{Hamming distance} $d(v,w)$ is defined as
   $$d(v,w)=\# \{i \in \{1, \ldots, n\} \ | \ v_i \neq w_i\}.$$
   The \textit{weight} $\omega(v)$ of the vector $v$ is its distance from the zero vector.
   Given a linear code $\C$, we say that $d$ is its \textit{minimum distance} if
    $$d=\min_{v \in \C \setminus \{0\}} \{\omega(v)\}.$$

Not all the combinations among the three parameters $[n,k,d]$ can be achieved, as claimed by the famous Singleton's bound, stating that we have the inequality $d+k \leq n+1.$

Now, we can introduce our main objects of study: Locally Recoverable Codes (LRCs). In what follows, given a subset $A \subset \{1, \ldots, n\}$, we denote by $\C_A$ the image of the code $\C$ with respect to the linear projection $\pi_A: \F_q^n \longrightarrow \F_q^{\mid A \mid}$. In coding theory, $\C_A$  is called the \emph{punctured} of $C$ at the complementary set of $A$, $\{1, \ldots, n\}\setminus A$

\begin{defn}\label{def LRC general}
     A code $\mathcal{C}$ of length $n$ and dimension $k$ is said to be $[r,\rho]$-\textit{locally recoverable} (LR), with $\rho \geq 2$, if every coordinate $i$ is contained in a subset $J_i \subset [n]$ of size at most $r+\rho-1$ such that the restriction $\C_{J_i}$ is a code of minimum distance $\rho$.
 \end{defn}

    In the context of the previous definition, the values of any $\rho-1$ coordinates of the recovery set $J_i$ are determined by the remaining $\#J_i - (\rho-1) \leq r$ symbols, allowing local recovery.

LRCs can be equipped with some additional properties, which are extremely useful for the recovery procedure: availability and hierarchical locality.

\begin{defn}\label{def availability LRC codes}
     A code $\mathcal{C}$ of length $n$ and dimension $k$ is said to be an LRC with \emph{availability} $t \geq 2$ ($LRC(t)$) if, for every coordinate $i$, the condition in \Cref{def LRC general} holds true for at least $t$ multiple subsets $\{J_{i,m}\}_{m=1, \ldots, t}$ such that $J_{i,m} \cap J_{i,m'}=\{i\}$ for every pair $(m,m')$. 
 \end{defn}

 \begin{defn}\label{def hierarchical LRC codes}
     A linear code $\C$ of length $n$ is said to have \emph{hierarchical locality} with parameters $([n_1,k_1,d_1]$, $[n_2,k_2,d_2])$ if for every $i \in \{1, \ldots, n\}$ there exists a subset $i \in J_i \subset [n]$ such that the punctured code $\C_{i,1}=\C_{J_i}$ has parameters $[n_1,k_1,d_1]$ and it is $[k_2,d_2]$-locally recoverable, with local recovery set $\C_{i,2}$ of length $n_2 \leq k_2+d_2-1$. The codes $\C_{i,1}$ and $\C_{i,2}$ are usually called respectively \textit{middle codes} and \textit{lower codes} for the $i$-th position.
 \end{defn}

 We conclude this brief subsection by presenting some known results that generalize the Singleton's bound for LRCs. 
  \begin{thm}[\scshape{Singleton's bound for LRCs}]
     The minimum distance of an $[r,\rho]$-locally recoverable code $\C$ of length $n$ and dimension $k$ satisfies the inequality:
     $$d \leq n-k+1 - \bigg(\left\lceil \frac{k}{r}-1 \right\rceil \bigg) (\rho-1).$$
     If the LRC has availability $t \geq 2$, two more general versions of the inequality hold true:
     $$d \leq n-k+2 - \left \lceil \frac{t(k-1)+1}{t(r-1)+1}\right \rceil;$$
     $$d \leq n-\sum_{i=0}^t \left \lfloor \frac{k-1}{r^i} \right \rfloor.$$
     If the $[n,k,d]$-code $\C$ is an HLRC with parameters $([n_1,k_1,d_1], [n_2,k_2,d_2])$, then
     $$d \leq n-k+1 - \bigg( \left \lceil\frac{k}{k_1} \right \rceil - 1 \bigg ) (d_1 - d_2) - \bigg( \left \lceil\frac{k}{k_2} \right \rceil - 1 \bigg ) (d_2 - 1).$$
 \end{thm}

We refer the reader to \cite[Appendix A]{kamath2013codeslocalregeneration},  \cite[Appendix A]{rawat2016}, \cite[\S IV.A]{tamo2016bounds} and \cite[\S II.B]{sasidharan2015} for the proofs of the bounds above.

\subsection{Algebraic and arithmetic geometry}

In this section, we recall some definitions and properties in algebraic geometry. We refer the reader to \cite{Hartshorne} for the theory of general varieties, to \cite{silverman2009} for elliptic curves, and to \cite{silverman2013advanced} for elliptic surfaces.
\smallskip

Given a field $\F$, an $\F$-\textit{variety} is an integral separated scheme of finite type over $\F$. We denote by $\F(X)$ its \textit{function field}. Given an $\F$-variety $X$, we will denote by $X(\F')$ the set of $\F'$-rational points of $X$, for any extension $\F'$ of $\F$.

Let $G$ be a finite group of automorphisms of $X$ and $Y$ be an $\F$-variety. A morphism $f: X \rightarrow Y$ of varieties is said to be \textit{$G$-invariant} if for every $g \in G$ we have $f \circ g = f$. A \textit{quotient} of $X$ by $G$ is a pair $(Y,\pi)$ such that:
\begin{itemize}
    \item $Y$ is a variety;
    \item $\pi: X \longrightarrow Y$ is $G$-invariant and, for every $G$-invariant morphism $f: X \longrightarrow Z$, there exists a unique morphism $\overline{f}: Y \longrightarrow Z$ such that $\overline{f} \circ \pi = f$.
\end{itemize}
If this quotient exists, we refer to it as $X/G$ instead of $Y$. For a complete construction of the quotient of a variety by a finite group of automorphisms, we refer the reader to \cite[\S 12.7]{gortz2020}.

\subsubsection{Curves}\label{sub: curves}

By a \textit{curve} we mean a smooth and geometrically irreducible $1$-dimensional variety over a field $\F$. 
In algebraic geometry, there exists a fundamental correspondence between geometric objects and algebraic objects. This relation allows us to study curves by working with their function fields. In fact, when dealing with quotients of curves, one can approach this topic by shifting from geometry to field extensions.

\begin{defn}
    Given a curve $X$ over a field $\F$, the group of divisors is the free abelian group $\DDiv(X)$ generated by the set of closed points $X^{(1)} \subseteq X$, \ie every divisor $D \in \DDiv(X)$ can be expressed as 
    $$D = \sum_{p \in X^{(1)}} n_p p,$$
    where the set $\{p \in X^{(1)} \ | \ n_p \neq 0\}$ is finite.
    For a fixed field extension $\F \subseteq \F'$, we denote by $\DDiv_{\F'}(X)$ the group of $\F'$-rational divisors on $X$, generated by the set of $\F'$-rational points.
\end{defn}

Usually, the set $X^{(1)}$ of closed points is also called the set of prime divisors of $X$. Moreover, one can associate a divisor to every rational function defined over a curve, as follows.

\begin{defn}
    Given an $\F$-curve $X$ and a rational function $f \in \F(X)^\ast$, its associated divisor is defined as
    $$\ddiv(f) = \sum_{p \in X^{(1)}} v_p(f) p,$$
    where $v_p$ is the discrete valuation defined over $\F(X)$ corresponding to the closed point $p \in X^{(1)}$.
\end{defn}

One can find a complete dissertation on discrete valuations over $\F(X)$ and the fact that the collection of points $\{p \in X^{(1)} \ | \ v_p(f) \neq 0\}$ is finite in \cite[Lem.~6.1]{Hartshorne}.

\begin{notat}
    Given an $\F$-curve $X$, let $D = \sum_{p \in X^{(1)}} n_p p$ be a divisor in $\Div(X)$.
    \begin{itemize}
        \item The support of $D$, denoted by $\Supp(D)$ is the set of prime divisors $p$ such that $n_p \neq 0$.
        \item If $n_p \geq 0$ for every $p \in X^{(1)}$, $D$ is said to be \textit{effective}.
        \item If there exists $f \in k(X)^\ast$ such that $D= \ddiv(f)$, $D$ is said to be a \textit{principal divisor}.
        \item A principal divisor can be expressed as
        $$\ddiv(f) = \ddiv_0(f) - \ddiv_\infty(f) = \sum_{v_p(f) > 0} v_p(f) p - \sum_{v_p(f) < 0} -v_p(f) p.$$
        If $p \in \Supp(\ddiv_0(f))$, the function $f$ is said to have a \textit{zero} of order $v_p(f)$ along $p$; if $p \in \Supp(\ddiv_\infty(f))$, the function $f$ is said to have a \textit{pole} of order $-v_p(f)$ along $p$. Thus, $\ddiv_0(f)$ and $\ddiv_\infty(f)$ are said to be respectively the divisor of zeros and the divisor of poles of $f$.
    \end{itemize}
\end{notat}

Maintaining the previous setup, the degree of a divisor $D = \sum_{p \in X^{(1)}} n_p p$ is the integer 
$$\deg(D) = \sum_{p \in \Supp(D)} n_p \cdot [\F(p):\F],$$ 
where $\F(p)$ is the smallest extension of $\F$ over which the closed point $p$ is defined. Thus, if $\Supp(D) \subseteq X(\F)$, then $\F(p) = \F$ for every $p \in \Supp(D)$ and $\deg(D) = \sum_{p \in \Supp(D)} n_p$. 

Two divisors $D_1,D_2 \in \Div(X)$ are said to be \textit{linearly equivalent} if there exists a rational function $f \in \F(X)^\ast$ such that $D_1 = D_2 + \ddiv(f)$. If $D_1$ and $D_2$ are linearly equivalent, we denote this relation by $D_1 \sim D_2$.

Given a divisor $D = \sum_{p \in X^{(1)}} n_p  p$ on a smooth curve $X$ defined over a field $\F$, we have a corresponding invertible sheaf $\OO_X(D)$ defined as follows on an open subset $\emptyset \neq U \subseteq X$:
        $$\OO_X(D)(U) = \{ f \in \F(U)=\F(X) \ | \ v_p(f) + n_p \geq 0 \ \forall p \in U^{(1)} \}.$$
The $\F$-vector space $H^0(X,\OO_X(D))$ of global sections of the invertible sheaf $\OO_X(D)$ is called the Riemann--Roch space associated to $D$. We usually denote by $h^0(D)$ its dimension over $\F$.

\begin{thm}[\scshape{Riemann-Roch on curves}, \protect{\cite[Ch.~IV, Thm.~1.3]{Hartshorne}}]
    Given a projective curve $X$ of genus $g$, let $D \in \DDiv(X)$. Then we have that:
    $$h^0(D) - h^0(K-D) = \deg (D) + 1 - g.$$
\end{thm}

The following known result will be useful for our code construction in \Cref{section: HLRC elliptic surfaces} and \Cref{section: HLRC with availability}. Since we cannot point to a reference, we give a proof for the sake of completeness.
\begin{prop}\label{prop: primitive_element_extension_group}
    Given an $\F$-curve $X$ and a finite group $G < \mathrm{Aut}(X)$, the field of rational functions $\F(X/G)$ is isomorphic to 
    $$\F(X)^G=\{f \in \F(X) \ | \ g \cdot f = f \ \forall \ g \in G\}.$$ Here, the action of an element $g \in G$ on a rational function $f \in \F(X)$ is defined as follows: for every point $p \in X$, $(g \cdot f)(p) = (f \circ g)(p)$.
    
    Moreover, if $m=|G|$ is coprime to $\mathrm{char}(\F)$, then the field extension $\F(X)/\F(X)^G$ is separable and has two additional useful properties:
    \begin{itemize}
        \item $[\F(X):\F(X)^G]=m$,
        \item there exists $f \in \F(X)$ such that
        $$\F(X)=\F(X)^G(f) \ \ \ \text{and} \ \ \ f \in H^0(X,\OO_X(D)),$$
        where $D$ is a divisor corresponding to a $G$-orbit in $X(\F)$.
        
    \end{itemize}
\end{prop}

\begin{myproof}
    The isomorphism between $\F(X/G)$ and $\F(X)^G$ can be proved as follows. Given any affine open subset $U=\Spec(A) \subseteq X$ invariant under the action of $G$, its quotient (\cite[\S 12.7]{gortz2020}) is defined as the open affine set $\Spec(A^G) \subseteq X/G$. Furthermore, $\F(X) = \mathrm{Frac}(A)$ and $\F(X/G) = \mathrm{Frac}(A^G)$. The isomorphism is proved if we show that $\mathrm{Frac}(A^G) = \mathrm{Frac}(A)^G$. In fact, the inclusion $\mathrm{Frac}(A^G) \subseteq \mathrm{Frac}(A)^G$ is trivial. Conversely, given a $G$-invariant element $f \in \mathrm{Frac}(A)$, it can be written as $f=\frac{a}{b}$, with $a,b \neq 0$ in $A$. It can also be expressed as
    $$f = \frac{a \prod_{g \in G \setminus \{1\}} g(b)}{b \prod_{g \in G \setminus \{1\}} g(b)} = \frac{a \prod_{g \in G \setminus \{1\}} g(b)}{\prod_{g \in G} g(b)}.$$
    Thus, $f$ and the denominator are both $G$-invariant: this implies that the numerator must be in $A^G$ and $f$ is contained in $\mathrm{Frac}(A^G)$.
    
    The field extension $\F(X)/\F(X)^G$ is a finite Galois extension, and we have that $[\F(X):\F(X)^G]=|G|=m$: we refer the reader to \cite[Ch.~VI, Thm.~1.8]{lang2012algebra} for these claims. For the second part of the statement, the existence of $f$ such that $\F(X)=\F(X)^G(f)$ is an immediate consequence of the primitive element theorem. Moreover, it is possible to give a characterization for the rational function $f$. Take $f \in \F(X)$ such that for every non-trivial $g \in G$, $g \cdot f \neq f$, \ie $f$ has trivial stabilizer in $G$. We know that the field extension $\F(X)/\F(X)^G$ is Galois with Galois group $G$, since it is normal and separable. If we show that 
    $$P_f(T) = \prod_{g \in G} (T- g \cdot f) \in \F(X)^G[T]$$
    is the minimal polynomial of $f$ over $\F(X)^G$, the proof is concluded: since the degree of $P_f(T)$ is exactly $m$, we would have that $[\F(X)^G(f):\F(X)^G]=m$, thus
    $$m=[\F(X):\F(X)^G] = [\F(X):\F(X)^G(f)] \cdot [\F(X)^G(f):\F(X)^G] = [\F(X):\F(X)^G(f)] \cdot m,$$
    which implies that $[\F(X):\F(X)^G(f)]=1$, that is equivalent to saying that $\F(X)=\F(X)^G(f)$. Every coefficient of the polynomial $P_f(T)$ is contained in $\F(X)^G$ because, for every $h \in G$, we have
    $$h \cdot P_f(T) = \prod_{g \in G} \big (T-h \cdot (g \cdot f) \big) = \prod_{g \in G} \big (T - (hg) \cdot f \big) = P_f(T).$$
    As a consequence, we have that the minimal polynomial of $f$ divides $P_f$. We just need to prove that $P_f$ is irreducible in $\F(X)^G$ among the polynomials that have $f$ as a root. By hypothesis $\mathrm{Stab}_G(f)$ is trivial and, from classical results in Galois theory, we have that
    $$\frac{|G|}{|\mathrm{Stab}_G(f)|}=[\F(X)^G(f):\F(X)^G].$$
    Thus, the minimal polynomial of $f$ over $\F(X)^G$ must have degree $m$, which is exactly the degree of $P_f$: this concludes the proof.

    In addition, one can prescribe the poles of this primitive element, such that they are contained in the support of a divisor $D$ on $X$ obtained from a $G$-orbit. Indeed, if $\{p_1, \ldots, p_m\}$ is a $G$-orbit, the Riemann--Roch space associated to the divisor $D=\sum p_i$ has dimension $$h^0(X,D) \geq m - g(X) + 1 \geq 1.$$
    For every non-trivial $g \in G$, the vector subspace 
    $$V_g = \{f \in H^0(X, \OO_X(D)) \ | \ g \cdot f = f \}$$
    is proper. If the field $\F$ is infinite, there exists $\overline{f} \in H^0(X, \OO_X(D)) \setminus \cup_{g \neq Id} V_g$ since a finite union of proper subspaces cannot cover the entire vector space. In fact, this is a primitive element of $\F(X)$ over $\F(X)^G$, since $\overline{f}$ has trivial stabilizer in $G$ and its poles are contained in the support of the divisor $D$. If $\F=\F_q$, the finite union $\cup_{g \neq Id} V_g$ may cover $H^0(X,\OO_X(D))$. In this case, one can notice that:
    $$\# H^0(X,\OO_X(D)) = q^{h^0(X,D)} \ \ \ \text{and} \ \ \ \# V_g \leq q^{h^0(X,D)-1}.$$
    Thus, $\# \cup_{g \neq Id} V_g \leq (m-1) \cdot q^{h^0(X,D)-1}$ and, if $q \geq m$, then 
    $$H^0(X,\OO_X(D)) \setminus \cup_{g \neq Id} V_g \neq \emptyset$$
    and we can conclude as above.    
\end{myproof}

Elliptic curves form the most studied classes of curves in algebraic geometry, number theory and complex analysis, with many real-world applications. In particular, these curves and their properties will be crucial in the next sections of this paper.
\begin{defn}\label{def: elliptic curve}
    An \textit{elliptic curve} $E$ over a field $\F$ is a projective curve of genus $1$ over $\F$, with a fixed point $p_0 \in E(\F)$.
\end{defn}
It is well-known that elliptic curves can be defined via a so-called long Weierstrass equation. Let $(E,p_0)$ be an elliptic curve over $\F$ and take $p_0 \in E(\F)$ as a fixed point. Then there exists a closed immersion $E \longrightarrow \p^2$ such that the image is the curve whose affine model is defined by an equation
$$y^2 + a_1 xy + a_3 y= x^3 + a_2x^2 + a_4x + a_6,$$
with $a_i \in \F$ such that the discriminant \footnote{The discriminant $\Delta(a_i)$ is defined as $\Delta(a_i)=-b_2^2b_8-8b_4^3-27b_6^2+9b_2b_4b_6$, such that $b_2 = a_1^2+4a_2$, $b_4=2a_4+a_1a_3$, $b_6=a_3^2+4a_6$, $b_8=a_1^2a_6+4a_2a_6-a_1a_3a_4+a_2a_3^2-a_4^2$.} $\Delta(a_i)$ is nonzero. Moreover, the point $p_0$ goes to the only point of the curve at infinity, which has classical coordinates given by $[0:1:0]$.

One can define a group structure on the set of points of an elliptic curve $E$. From now on, the fixed point $p_0 \in E(\F)$ will be denoted by $\infty$.

\begin{constr}
    Given an elliptic curve $(E,\infty)$ defined over a field $\F$, define the map
    \[
    \funcDef{\Phi}{E(\F)}{{\Pic}^0(E)}
    {p}{\OO_E(p-\infty)},
    \]
    where $\Pic^0(E)$ denotes the subgroup of $\Pic(E)$ which consists of the classes of invertible sheaves corresponding to divisors of degree $0$. This map gives a bijective correspondence between the set of points $E(\F)$ and the elements of the group $\Pic^0(E)$, which furnishes us with a group structure on $E(\F)$, with $\infty$ as identity and operation denoted by $\oplus$.
\end{constr}

\begin{notat}
    Given a divisor $D= \sum n_p \cdot p$ on an elliptic curve $E/\F$ such that $\Supp(D) \subseteq E(\F)$, we use the following notation:
    $$\sum^\oplus D = \sum^\oplus p^{\oplus n_p} \in E(\F).$$
    In fact, $\sum^\oplus D$ is the point on $E$ obtained as the sum, with multiplicity, of the points in the support of the divisor $D$, with respect to the operation of the group law on $E(\F)$. Moreover, if $G < E(\F)$ is a subgroup of points, we denote by $D_G \in \DDiv_{\F}(E)$ the divisor defined by the formal sum of the points contained in the subgroup $G$.
\end{notat}

We now recall the definition of isogenies and torsion points.

\begin{defn}
    Given two elliptic curves $(E_1,\infty_1)$ and $(E_2,\infty_2)$ over $\F$, an \textit{isogeny} is a non-constant morphism of $\F$-schemes $f: E_1 \longrightarrow E_2$ such that $f(\infty_1)=\infty_2$.
\end{defn}

We have two relevant examples of isogenies.

\begin{ex}
    Given an elliptic curve $E$ defined over $\F=\overline{\F}$ and an integer $m \in \Z \setminus \{0\}$, the multiplication-by-$m$ homomorphism
    \[
    \funcDef{[m]}{E}{E}
    {p}{p^{\oplus m}}
    \]
    is an isogeny.
\end{ex}

\begin{ex}
    Another source of isogenies between elliptic curves is given by the quotients with respect to a finite subgroup $G < E(\F)$ of order which is not divisible by $\mathrm{char}(\F)$. Indeed, $G$ can be seen as a subgroup of $\mathrm{Aut}(E)$, since every point $p \in G$ defines the translation
    \[
    \funcDef{\tau_p}{E}{E}
    {q}{q\oplus p}
    \]
    The quotient $(E/G,\pi)$ is still an elliptic curve and every point $q \in E/G(\F)$ has exactly $|G|$ points in the pre-image. One can conclude that $\pi: E \longrightarrow E/G$ is an isogeny between elliptic curves because $\pi(\infty_E) = \infty_{E/G}$ and $\pi$ is clearly non-constant.
\end{ex}

The first example leads us to the formal definition of torsion points. The second example, associated with quotients of elliptic curves, will be crucial for the construction in \Cref{section: HLRC elliptic surfaces} and \Cref{section: HLRC with availability}.

\begin{defn}
    Given an elliptic curve $(E,\infty)$ defined over a field $\F$ and an integer $m \in \Z \setminus \{0\}$, 
    $$\mathrm{ker}[m]=\{p \in E(\F) \ | \ [m]p=\infty\}$$
    is called \textit{$m$-torsion subgroup} of $E(\F)$ and usually denoted by $E(\F)[m]$.
\end{defn}

\begin{prop}\label{prop: torsion subgroups all results}
    Let $(E,\infty)$ be an elliptic curve defined over a field $\F$ such that $\mathrm{char}(\F)=p \geq 0$ and consider an integer $m \in \Z \setminus \{0\}$. Given a fixed algebraic closure $\overline{\F}$, we have
    \begin{itemize}
        \item[$\mathbf{(a)}$] if $\mathrm{gcd}(p,m)=1$, $E(\overline{\F})[m] \cong \Z/m \times \Z/m$;
        \item[$\mathbf{(b)}$] if $p > 0$, we have two possibilities:
        \begin{itemize}
            \item $E(\overline{\F})[p^r] \cong \Z/p^r$ for every $r \geq 1$;
            \item $E(\overline{\F})[p^r] = \{\infty\}$ for every $r \geq 1$;
        \end{itemize}
        \item[$\mathbf{(c)}$] if $p > 0$, $q=p^n$, the $\F_q$-rational points define a group of the form
        $$E(\F_q) \cong \Z/a \times \Z/b,$$
        with $a | b$ and $\mathrm{gcd}(a,p)=1$.
    \end{itemize}
\end{prop}

The last result of this subsection is Abel's Theorem, which will be crucial in our first construction.

\begin{thm}[\scshape{Abel}]\label{thm: abel}
     Let $E/\F$ be an elliptic curve, whose group of points is defined as $(E(\F),\infty)$. Let $D=\sum n_p \cdot p$ be a divisor in $\DDiv_\F(E)$. Then 
    $$D \sim \bigg(\sum^\oplus D\bigg) + (\deg(D)-1)\cdot \infty.$$
    In particular, $D$ is a principal divisor if and only if
    $$\sum^\oplus D = \infty \in E(\F) \ \ \ \text{and} \ \ \ \deg(D)=0.$$ 
\end{thm}

\begin{myproof}
    We refer the reader to \cite[Ch.~III, Prop.~3.4 and Cor.~3.5]{silverman2009} for a proof of the second part of the statement. The first claim on a divisor $D \in \DDiv_\F(E)$ can be proved by considering the degree zero divisor $D' = D - \deg(D) \cdot \infty$. 
\end{myproof}

\subsubsection{Surfaces}

By a \textit{surface} we mean a smooth and geometrically irreducible $2$-dimensional variety over $\F$. 
In this paper, we will work on a specific class of surfaces, that is, elliptic surfaces.

\begin{defn}
        Let $B$ be a projective curve. An \textit{elliptic surface} $\E$ over $B$ consists of the following data:
        \begin{itemize}
            \item a projective $\F$-surface $\E$;
            \item a morphism $\pi: \E \longrightarrow B$ such that for all but finitely many $b \in B(\overline{\F})$, the fiber $E_b = \pi^{-1}(b)=\E \times_B \Spec \ \F(b)$ is a smooth projective curve of genus $1$;
            \item a section $\sigma_0: B \longrightarrow \E$, \ie a morphism $\sigma_0$ such that $\pi \circ \sigma_0 = Id_B$.
        \end{itemize}
    \end{defn}

\begin{rem}
    The section $\sigma_0$ of the definition above fixes an $\F$-rational point on every fiber of the morphism $\pi$. Thus, given $b \in B$ such that $E_b$ is a smooth projective curve of genus $1$, it is in fact an elliptic curve according to Definition \ref{def: elliptic curve}.
\end{rem}

    There exists a correspondence between elliptic $\F$-surfaces $\E$ over a curve $B$ and elliptic curves $E$ over the function field $\F(B)$, as stated in the following proposition \cite[Cha.~III, Proposition~3.8]{silverman2013advanced}.
    \begin{prop}\label{prop_corr_ell_curves_surfaces}
    Given a curve $B$ over $\F$, let $E$ be an elliptic curve defined over the field $\F(B)$, with long Weierstrass equation $y^2+a_1 xy + a_3y = x^3 + a_2x^2 + a_4 x + a_6$ such that $a_i \in \F(B)$. We have a corresponding elliptic surface $\E(a_i) \subseteq \p^2_B = \p^2_\F \times_\F B$ defined by the equation 
    $$Y^2Z+a_1XYZ + a_3YZ^2=X^3+a_2X^2Z + a_4XZ^2 + a_6Z^3,$$
    where the fibration is given by the projection to the second factor of $\p^2_\F \times_\F B$. 
            
    Given an elliptic surface $\pi: \E \longrightarrow B$ defined over $\F$, there exist $a_i \in \F(B)$ such that $\E$ is $\F$-birationally equivalent to $\E(a_i)$ defined above. Moreover, the elliptic curve $E/\F(B)$ defined by the equation $y^2+a_1xy+a_3y=x^3+a_2x^2+a_4x+a_6$ is uniquely determined by $\E$ up to $\F(B)$-isomorphisms.
    
    Given an elliptic surface $\E \longrightarrow B$ associated to an elliptic curve $E/\F(B)$ as above, we have the following isomorphism of $\F(B)$-algebras:
    $$\F(\E) \cong \F(B)(E).$$
\end{prop}

\begin{rem}
    Given an elliptic $\F$-surface $\pi: \E \longrightarrow B$, the corresponding elliptic curve $E$ defined over the field $\F(B)$ is the fiber over the generic point $\eta \in B$. In fact, this curve is called the generic fiber of $\E$ over $B$ and the set of $\F(B)$-rational points of $E$ is in bijection with the set of the global sections of the surface, \ie the set of morphisms $\sigma: B \longrightarrow \E$ such that $\pi \circ \sigma = \mathrm{Id}_B$.
\end{rem}

In \Cref{section: HLRC elliptic surfaces} and \Cref{section: HLRC with availability}, we will use the generic fiber to pass from a construction on an elliptic curve to one on an elliptic surface. For this purpose, a crucial algebro-geometric instrument is the specialization map, defined from the generic fiber to another elliptic fiber.

Consider an elliptic $\F$-surface $\pi: \E \longrightarrow B$ and its generic fiber $E/\F(B)$. Let $\gamma \in B(\F)$ be a point such that the corresponding fiber $E_\gamma = \E \times_B \Spec (\F(\gamma))= \E \times_B \Spec (\F)$ is an elliptic curve. We can define a specialization map $\Sp_\gamma: E(\F(B)) \longrightarrow E_\gamma(\F)$ as follows. A point $p$ of $E(\F(B))$ is a global section $\sigma_p: B \longrightarrow \E$, while $\gamma$ corresponds to a morphism $i_\gamma: \Spec(\F) \longrightarrow B$. Moreover, we have also the two projections $\pi_1$ and $\pi_2$, which fit in the commutative diagram below:

\begin{equation} \label{diag:universal_specialization}
\begin{tikzcd}[row sep=large, column sep=large]
\text{Spec}(\mathbb{F}) \arrow[drr, "\sigma_P \circ i_\gamma", bend left=20] \arrow[ddr, "\text{id}"', bend right=20] \arrow[dr, "\exists ! \, P_\gamma" description, dashed] & & \\
& E_\gamma \arrow[r, "\pi_1"] \arrow[d, "\pi_2"'] & \mathcal{E} \arrow[d, "\pi"] \\
& \text{Spec}(\mathbb{F}) \arrow[r, "i_\gamma"'] & B
\end{tikzcd}
\end{equation}
By the universal property of the fibered product, there exists a unique morphism $p_\gamma: \Spec(\F) \longrightarrow E_\gamma$, which in fact is an $\F$-rational point of the elliptic fiber $E_\gamma$.
\begin{defn}
    In the setting described above, the specialization map at $\gamma \in B(\F)$ is the function
    \[
    \funcDef{\Sp_\gamma}{E(\F(B))}{E_\gamma(\F)}
    {\sigma_p}{p_\gamma}.
    \]
\end{defn}

\begin{thm}\label{thm: specialization map}
    Given the above setup, the set-theoretic specialization map
    \[\funcDef{\Sp_\gamma}{E(\F(B))}{E_\gamma(\F)}
    {\sigma_p}{p_\gamma}
    \]
    is a surjective homomorphism of groups. Moreover, if $m$ is a positive integer coprime with the characteristic of the field $\F$, then the map
    $$E(\F(B))[m] \longrightarrow E_\gamma(\F)$$
    is injective.
\end{thm}

\begin{myproof}
This theorem can be proved relying on the standard theory of reduction for elliptic curves over local fields: indeed, a non--singular fiber $E_\gamma$ in our setting corresponds to an elliptic curve $E$ with good reduction at a place $\gamma$. Analogously, the specialization map $\mathrm{sp}_\gamma$ corresponds to a reduction map. Thus, the theorem is a consequence of Propositions $2.1$, $2.2$ and Theorem $3.1$ contained in \cite[Ch.~VII]{silverman2009}.   
\end{myproof}

\begin{rem}\label{rem: equivalence evaluation}
	Given the above setup, let $K$ be the field $\F(B)$ and consider a function $f \in K(E) = \F(\E)$, where $E/K$ is the generic fiber of $\E$ and take $p \in E(K)$ and $\gamma \in B(\F)$ such that $f \in \OO_{E,p} \subset K(E)$ and $f(p) \in \OO_{B,\gamma} \subset \F(B)$. We want to prove that:
	$$f(p)(\gamma) = f_\gamma(p_\gamma),$$
	where $p_\gamma=\text{sp}_\gamma(p)$ and $f_\gamma = f_{|E_\gamma} = f \circ j_\gamma$, with $j_\gamma: E_\gamma \longrightarrow \E$ inclusion. We denote by $\mathcal{M}_p$ the maximal ideal of $\OO_{E,p}$ and by $\mathcal{M}_\gamma$ the maximal ideal of $\OO_{B,\gamma}$. By definition, 
	$$f(p)(\gamma) = \big[\overline{f_p}\big] \in \OO_{B,\gamma}/\mathcal{M}_\gamma,$$
	where $\overline{f_p}$ is the class of $f$ in $\OO_{B,\gamma} \subset \OO_{E,p}/\mathcal{M}_p = K$. In fact, we are taking the class of $f$ modulo $\mathcal{M}_p$ and modulo $\mathcal{M}_\gamma$.
	
	On the other hand, taking the restricted function $f_\gamma$ is equivalent to considering the quotient of $f$ with respect to the ideal of the closed fiber $E_\gamma$, which is clearly generated by the local parameter $t_\gamma$ at $\gamma$. After that, evaluating $f_\gamma$ at $p_\gamma$ corresponds to taking its class modulo $\mathcal{M}_\gamma$. 
	
	Thus, geometrically both sequences of quotients evaluate $f$ at the intersection point between the section $\sigma_p$ and the closed fiber $E_\gamma$. Algebraically, both sequences correspond to the quotient with respect to the maximal ideal of the aforementioned intersection point.
	
\end{rem}

 \subsection{Algebraic Geometry codes}

 Algebraic Geometry codes (AG codes) were introduced by Goppa in \cite{goppa1983} over smooth absolutely irreducible curves $X$ defined over a finite field $\F_q$. His idea was to evaluate the rational functions contained in a certain vector subspace of $\F_q(X)$ over a given subset of the rational points $X(\F_q)$. Goppa's construction was successively extended to projective varieties of arbitrary dimension, as presented below.
 \begin{defn}\label{def_ag_codes} Consider an absolutely irreducible smooth projective variety $X$ defined over the finite field $\F_q$ and a set $T=\{p_1, \ldots, p_n\} \subseteq X(\F_q)$. Let $V \subseteq \F_q(X)$ be a vector space of finite dimension, consisting of rational functions such that the supports of their divisors of poles avoid the collection of points $T$. 
    The AG code $\mathcal{C}(X,V,T)$ is defined as the image of the following $\F_q$-linear evaluation map
    \[
    \funcDef{\ev_T}{V}{\F_q^n}
    {f}{(f(p_1), \ldots, f(p_n))}.
    \]
\end{defn}

One can choose the vector space $V$ as a specific Riemann-Roch space $H^0(X, \OO_X(G))$ associated to a divisor $G \in \Div_{\F_q}(X)$ such that its support does not contain any of the points of $T$. In this case, if there is no ambiguity regarding the divisor $G$, the corresponding AG code is denoted by $\C(X,G,T)$. In fact, the hypothesis on the divisor $G$ avoiding the set of rational points $T$ is not necessary: indeed, we have that $\OO_X(G)$ is an invertible sheaf, so for every point $p \in X$ the stalk of $\OO_X(G)$ at $p$ is an $\OO_{X,p}$-module generated by $s_p$. Then any global section $f$ can be locally written as $f_p s_p$ for some $f_p \in \OO_{X,p}$, which one can evaluate at the point $p$. Thus, we can define any AG code over a variety of arbitrary dimensions on the whole set of rational points. 

To end this section, we give a survey of the known results about the parameters of AG codes from curves, which we shall use in the paper. 

\begin{thm}\label{prop: AG codes curves}
    Let $X$ be a curve over $\F_q$, $G \in \Div(X)$, $T=\{p_1, \ldots, p_n\} \subseteq X(\F_q)$ and $D=\sum_{p \in T} p$ the associated divisor. Then, the parameters $[n,k,d]$ of the code $\mathcal{C}(X,G,T)$ satisfy:
    \begin{itemize}
        \item $n=\#T$,
        \item $k=h^0(G) - h^0(G-D)$. If $n=\deg(D) > \deg(G)$, then $k=h^0(G)$,
        \item $d \geq n-\deg(G)$.
    \end{itemize}
\end{thm}

One can consult \cite[\S 4.1.1]{AGBook} for a proof of the theorem above. There exist various analog estimates for the parameters of AG codes on surfaces---see, for example, \cite{ABHP21b,CouvreurEtAl2020}. These often make use of intersection theory on surfaces and go beyond our interests, since the constructions contained in the following sections strongly rely on curves.

\section{Availability from elliptic surfaces}\label{section: availability ell surfaces}

In this section, we are going to define our locally recoverable code on the elliptic surface $\pi: \mathcal{E} \longrightarrow B$ defined over $\F_q$.  We take inspiration from the construction in \cite{SVV}, but slightly modify the key lemma in it to obtain several disjoint recovery sets for every point in the evaluation set, thus LRC with availability.

The crucial point of this construction lies in the choice of the evaluation set for the code. Consider a subset $S \subset B(\F_q)$ such that for every element $\gamma \in S$, the fiber $\pi^{-1}(\gamma)$ is an elliptic curve. Take now a prime number $m > 3$, different from the characteristic $p$ of the field. The subgroup of $m$-torsion of an elliptic curve $E_\gamma$ defined over the algebraically closed field $\overline{\F}_q$ is isomorphic to $\Z/m \times \Z/m$.

In this section, we will work over the finite extension $\F$ of $\F_q$ over which the $m$-torsion subgroup of every elliptic curve $E_\gamma$, $\gamma \in S$, is defined. We will denote by $G_\gamma$ the $m$-torsion subgroup of $E_\gamma$, and by $T_\gamma$ the subset $G_\gamma \setminus \{\infty_\gamma\}$, consisting of exactly $m^2-1$ distinct points. 

Then, the evaluation set for our construction is defined as 
$$T = \bigcup_{\gamma \in S} T_\gamma = \bigcup_{\gamma \in S} \big(G_\gamma \setminus \{\infty_\gamma\}\big).$$

Given $\gamma \in S$, the group $G_\gamma$ can be seen as the linear space $\F_m^2$ over $\Z/m=\F_m$ of dimension $2$.  Thus, if we take a point $p_0 \in G_\gamma$, exactly $m+1$ lines are passing through $p_0$, corresponding to the possible slopes, which, in fact, are the elements of $\p^1_{\F_m}(\F_m)$. 
One can also prove that any two of these lines intersect only at the point $p_0$. Let $L_{\gamma,p_0,i}$ (or simply $L_{\gamma,i}$ if no confusion arises on the point $p_0$) be one of the lines in $(\F_m)^2$ passing through $p_0$, where $i=1, \ldots, m+1$. We index this collection of lines such that $L_{\gamma,p_0,m+1}$ is the line passing through $p_0$ and $\infty_\gamma$ in $(\F_m)^2$. Finally, for every $i \in \{1, \ldots, m\}$, we let $T_{\gamma,p_0,i}$ denote the set of $m$ points contained in the support of the line $L_{\gamma,p_0,i}$.

\begin{lem}\label{lem:sumtoinf}
    Given $\gamma \in S$, $m$ an odd prime different from $p=\mathrm{char}(\F)$ and a point $p_0 \in T_\gamma$, the sets $T_{\gamma,p_0,i} \subset T_\gamma$ satisfy the following condition, for every $i\in\{1, \ldots, m\}$:
    $$\sum_{p \in T_{\gamma,p_0,i}}^\oplus p = \infty_\gamma.$$
    Moreover, none of the sets $T_{\gamma,p_0,i}$, with $i \in \{1, \ldots, m\}$, contains the point $\infty_\gamma$. 
\end{lem}

\begin{myproof}
    The second claim follows immediately from the fact that there exists only one line passing through $p_0$ and $\infty_\gamma$, and we fixed this line to be $L_{\gamma,p_0,m+1}$. For the main part of the proof, we are going to work with the group $\Z/m \times \Z/m$ isomorphic to $G_\gamma$. We will prove that the sum of the points contained in the support of a line in $\Z/m \times \Z/m = (\F_m)^2$ gives the point $(0,0)$, which corresponds to the point $\infty_\gamma \in G_\gamma$. 
    
    Every line $L$ in $(\F_m)^2$ can be parametrized considering its direction $v= (a,b)$ and a point $(x_0,y_0)$ contained in its support. In particular, any point $p \in \Supp(L)$ is then of the form
    $$p=p_t = (x_0,y_0) + (ta,tb), \ \ \ \text{for some } \ t \in \{0, \ldots, m-1\}.$$
    Thus, the sum of the points in $\Supp(L)$ corresponds to the sum 
    \begin{align*}
        \sum_{t}^{\oplus} p_t &= \bigg(\sum_{t=0}^{m-1} (x_0+ta),\sum_{t=0}^{m-1} (y_0+tb)  \bigg)=\bigg(mx_0 + a\sum_{t=0}^{m-1} t, my_0 + b\sum_{t=0}^{m-1} t\bigg) \\
        &= \bigg( mx_0 + \frac{m(m-1)}{2} a, my_0 + \frac{m(m-1)}{2} b \bigg).
    \end{align*}
    By hypothesis, $m$ is an odd prime number, so $\frac{m-1}{2}$ is an integer and both coordinates of the point defined above are divisible by $m$, thus, it is the point $(0,0)$, which corresponds to the point $\infty_\gamma \in G_\gamma$. This concludes the proof of the lemma.
\end{myproof}

This lemma furnishes us with the necessary assumptions for the complete construction of the LRC with availability $t=m$ that will be explained in the following proposition. Recall that given a prime number $m > 3$ different from the characteristic of the field $\F$, $\E$ is an elliptic $\F$-surface, with a fibration $\pi: \E \longrightarrow B$ such that $E_\gamma[m] \subseteq E_\gamma(\F)$ for every $\gamma \in S \subset B(\F_q)$, defined as in the first paragraph of this section. 

\begin{thm}\label{prop: hier_availability_ell_surfaces}
   Consider a vector space $V \subseteq \F(\mathcal{E})$ such that for every $\gamma \in S$ the image of the restriction  
    \[
    \funcDef{\mathrm{res}_{E_\gamma}}{V}{\F(E_\gamma)}
    {f}{f_{|E_\gamma}}
    \]
   is exactly $H^0(E_\gamma,\OO_{E_\gamma}((m-1)\infty_\gamma))$. The evaluation code $\mathcal{C}(\mathcal{E},V,T)$ obtained evaluating the rational functions of $V$ on the points of
   $$T=\bigcup_{\gamma \in S} T_\gamma = \bigcup_{\gamma \in S} \bigg(G_\gamma \setminus \{\infty_\gamma\}\bigg)$$
   is locally recoverable, with locality $r=m-1$ and availability $t=m$. In particular, every point $p_0 \in T_\gamma$ has $m$ disjoint recovery sets $T_{\gamma,p_0,i}$, with $i= \{1, \ldots, m\}$. This means that the evaluation of a rational function at a point $p_0 \in T_\gamma$ can be recovered using the points in the support of one of the lines passing through $p_0$, except for the line through $(0,0)$ in $(\F_m)^2$.
\end{thm}

\begin{myproof}
    The proof of this proposition is largely inspired by that of \cite[Lem.~VI.1]{SVV}. By definition, for every point $p_0 \in T$ there exists a unique $\gamma \in S$ such that $p_0 \in T_\gamma$. Given a rational function $f \in V$ and a point $p_0 \in T_\gamma$ such that $f(p_0)$ is unknown, we want to prove that this value can be recovered using the evaluations of $f$ at the $r=m-1$ points contained in $T_{\gamma,p_0,i}\setminus \{p_0\}$ and this for every $i \in \{1, \ldots, m\}$. This is equivalent to proving that the following $\F$-linear evaluation map 
    \[
            \funcDef{\ev_{T_{\gamma,i} \setminus \{p_0\}}}{H^0(E_\gamma,\OO_{E_\gamma}(r\infty_\gamma))}{\F^r}
            {f}{(f(p))_{p \in T_{\gamma,i} \setminus \{p_0\}}}
       \]
    is an isomorphism for every $i \in \{1, \ldots, m\}$. Applying the Riemann--Roch theorem, we know that $h^0(E_\gamma,\OO_{E_\gamma}(r\infty_\gamma))=\mathrm{deg}(r\infty_\gamma)=r$, so we just need to prove that the map above is injective.

    Suppose that two distinct rational functions $f_1$ and $f_2$ assume the same values at the points of $T_{\gamma,i} \setminus \{p_0\}$. Then, the function $g=f_1-f_2$ has an associated divisor given by
    $$\ddiv(g) = \sum_{p \in T_{\gamma,i} \setminus \{p_0\}} p - r \infty_\gamma.$$
    Applying Abel's theorem, we obtain 
    $$\sum_{p \in T_{\gamma,i} \setminus \{p_0\}}^\oplus p = \infty_\gamma.$$
    This implies that $$\sum_{p \in T_{\gamma,i}}^\oplus p= p_0 \oplus \sum_{p \in T_{\gamma,i} \setminus \{p_0\}}^\oplus p = p_0 \oplus \infty_\gamma = p_0,$$ which contradicts \Cref{lem:sumtoinf}. Thus, the evaluation map above is an isomorphism and the code is locally recoverable, with locality $r=m-1$ and availability $t=m$.

\end{myproof}

\begin{rem}
    Above, we have proved that for every $p_0 \in T_\gamma$, the restriction of the evaluation map to the points contained in $T_{\gamma,1,j_i} \setminus \{p_0\}$ is an isomorphism, but we have not mentioned the recovery procedure, which we explain here. If $E_\gamma$ has long Weierstrass equation $y^2+a_1xy+a_3y=x^3+a_2x^2+a_4x+a_6$, the Riemann--Roch space $H^0(E_\gamma,\OO_{E_\gamma}(r\infty_\gamma))$ is generated by 
    $$\{x^i \cdot y^j \ | \ 2i+3j \leq r \ \text{and} \ j \in \{0,1\}\}.$$
    Then, we can see it as an interpolation problem and use the classic bivariate interpolation to recover the missing value.
\end{rem}

To conclude this section, we give in the following theorem the parameters of the code we have constructed above.

\begin{prop} \label{prop: parameters_LRC(2)_ell_surfaces}
    We let the hypotheses and notation for the integers $m$ and $r$, the sets $S$ and $T$, and the field $\F$, be as above. For $B$ a projective curve of genus $g$, consider the elliptic surface $\pi: \E \longrightarrow B$, with Weierstrass equation
    $$y^2 +a_1xy + a_3y = x^3 + a_2x^2 + a_4 x+ a_6, \ \ \ \text{such that} \ a_i \in \F(B).$$ Let $V$ be the $\F$-vector space defined as 
    $$V= \bigg \{ \sum_{\substack{2i+3j \le r \\ j \in \{0,1\}}} b_{i,j} x^i y^j \ | \ b_{i,j} \in H^0(B,\OO_B(D)) \bigg \},$$
    where $D \in \DDiv_{\F}(B)$ is an effective divisor  of degree $\delta$, with $g \leq \delta < N = \#S$, $\delta>1$, and such that $S \cap \Supp(D) = \emptyset$. Let $\mathcal{C}(\mathcal{E},V,T)$ be the code obtained by evaluating the rational functions of $V$ on the points of $T$.
    Then, the parameters $[n,k,d]$ of $\mathcal{C}(\mathcal{E},V,T)$ respect the following bounds:
    \begin{itemize}
        \item[$\mathbf{(a)}$] $n= N \cdot (m^2-1) = N \cdot (r^2+2r)$;
        \item[$\mathbf{(b)}$] $k \geq (\delta - g +1) \cdot (m-1) = (\delta - g +1) \cdot r$;
        \item[$\mathbf{(c)}$] $d \geq (N-\delta) \cdot \bigg [(m^2-1) - (m-1)\bigg ] = (N-\delta) \cdot (m^2-m) = (N-\delta) \cdot (r^2+r)$.
    \end{itemize}
\end{prop}

    \begin{myproof}
        The length $n$ of the code can be computed by counting the number of points contained in each subset $T_\gamma \subseteq E_\gamma(\F)$, with $\gamma \in S$: since $T_\gamma = G_\gamma \setminus \{\infty_\gamma\}$, the cardinality of every $T_\gamma$ is $m^2 - 1$. Thus, the length of the full code is exactly $\#S \cdot (m^2-1) = N \cdot (m^2-1)$, as claimed.

        The idea to prove the statement on the minimum distance is the following: we want to count the maximum number of elliptic curves $E_\gamma$ over which a non-trivial function $f \in V$ can be identically zero; then we count the maximum number of zeros that can occur on a single elliptic curve. If we take a non-trivial function $f \in V$, it can be expressed as
        $$f= \sum_{\substack{2i+3j \le r \\ j \in \{0,1\}}} b_{i,j} x^i y^j,$$
        with $b_{i,j} \in H^0(B,\OO_B(D))$ for every $i,j$. If we fix a point $\gamma \in S$, the restriction $f_{|E_\gamma}$ is a polynomial in $x,y$ with coefficients $b_{i,j}(\gamma)$. Thus, $f$ is identically zero on $E_\gamma$ if and only if $b_{i,j}(\gamma)=0$ for every $i,j$, \ie $\gamma$ is a common zero of the sections in the collection $\{b_{i,j}\} \subseteq H^0(B,\OO_B(D))$. Given $i_0,j_0$ such that $2i_0 + 3j_0 \leq r$ and $j_0 \in \{0,1\}$, the set of common zeros of all these functions is a subset of $\Supp(\ddiv_0(b_{i_0,j_0}))$. The support of this divisor of zeros has cardinality $\delta' \leq \delta$ (it is not necessarily an equality since we are not counting multiplicities in this set). Therefore, a non-trivial $f \in V$ can be identically zero on at most $\delta$ fibers.
        
        For the last part of the proof, we are going to count the maximum number of zeros of a non-trivial function on an elliptic curve using the fact that every rational function $f \in V_\gamma$ is contained in $H^0(E_\gamma, \OO_{E_\gamma}((m-1)\infty_\gamma))$. As such, the number of zeros counted with multiplicity, which is equal to the number of poles counted with multiplicity, is at most $m-1$. Combining these two results, we obtain that there are at least $N-\delta > 0$ elliptic curves of the form $E_\gamma$ over which a non-trivial function $f \in V$ is not identically zero. Moreover, on each of these curves, $f$ has at most $m-1$ zeros. Then, the evaluation map is injective since $m-1 < m^2-1$, which is the cardinality of every set $T_\gamma$, and the minimum distance is at least
        $$(N-\delta) \cdot \bigg [(m^2-1) - (m-1)\bigg] = (N-\delta) \cdot (m^2-m) = (N-\delta) \cdot (r^2+r).$$

        Finally, since the evaluation map is injective, the dimension $k$ of the code is equal to the $\F$-dimension of the vector space $V$. We have
        $$k = \dim_\F V = h^0(B,\OO_B(D)) \cdot r \geq (\delta-g+1) \cdot r.$$
        The second equality follows from the definition of the vector space $V$ and the fact that for every elliptic curve $E_\gamma$, $h^0(E_\gamma,r\infty_\gamma)=r$ by the Riemann--Roch theorem. Finally, applying the Riemann--Roch theorem, we obtain that $h^0(B,\OO_B(D)) \geq \delta-g+1$.

    \end{myproof}

\section{Hierarchical locality from elliptic curves and surfaces}\label{section: HLRC elliptic surfaces}

This section aims to construct hierarchical LRCs (HLRCs) from elliptic surfaces. We will use some well-chosen quotients of elliptic curves and exploit results exposed in \cite[\S IV]{ballentine2019}. In the latter paper, the authors worked on arbitrary curves and separable morphisms between them. By restricting our focus to the specific case of elliptic curves, we refine their results into a systematic approach for the selection of the evaluation set and the rational functions to be evaluated at it. One can notice that a hierarchical structure can be achieved on individual elliptic curves, while the transition to elliptic surfaces adds complexity to the hierarchical structure: the fibers over a base curve are naturally elliptic curves, thus our construction acquires an additional layer. The structure of the code will be made explicit in the subsequent paragraphs. 

\subsection{HLRCs from a sequence of quotients of elliptic curves}\label{subsection: hlrc quotients elliptic curves}

In this section, we consider an elliptic curve $E$ defined over a field $\F_q$ such that $E[m] \subseteq E(\F_q)$, for a fixed integer $m$ such that $3 < m < q$ and $\gcd(m,q)=1$. Note that in contrast with the previous section, we no longer impose $m$ to be prime. These are only our initial assumptions on the integer $q$. Indeed, later we will impose further conditions on the size of $q$: we want it to be sufficiently large to guarantee the existence of rational functions with suitable properties for the interpolation procedure.

Following the construction given in \cite[\S IV]{ballentine2019}, we want to define a sequence of separable morphisms starting from the curve $E$. Consider two independent points $m_1$ and $m_2$ in $E[m] \subseteq E(\F_q)$ of order exactly $m$. We know that 
$$G_1=\langle m_1 \rangle \cong  \langle m_2 \rangle \cong \Z/m \ \ \ \text{and} \ \ \ \langle m_1, \ m_2 \rangle = E[m].$$
The quotient isogeny
$$\varphi_1: E \longrightarrow E_1 = E/G_1$$
is defined over $\F_q$, has kernel $G_1$ and degree $m$. Let $\overline{m_2}$ denote the image $\varphi_1(m_2) \in E_1(\F_q)$. The point $\overline{m_2}$ is indeed $\F_q$-rational because $G_1 < E(\F_q)$, the elliptic curve $E/G_1$ and the quotient isogeny are defined over $\F_q$. Since $m_1$ and $m_2$ are independent and $m_2$ has order $m$, the point $\overline{m_2}$ has order $m$ in $E_1(\F_q)$. Let us define $G_2 = \langle \overline{m_2} \rangle \subseteq E_1[m]$. Then, the quotient isogeny 
$$\varphi_2: E_1 \longrightarrow E_2 = E_1/G_2 = E/E[m]$$
has degree $m$. Thus, we have obtained a sequence of morphisms 
$$E \stackrel{\varphi_1} \longrightarrow E_1 \stackrel{\varphi_2} \longrightarrow E_2 = E/E[m]$$
such that the composition $\psi = \varphi_2 \circ \varphi_1$ is an isogeny defined over $\F_q$ with kernel given by $\langle m_1, m_2 \rangle = E[m]$. All the morphisms considered so far are separable because $m$ is coprime with the characteristic of the field $\F_q$.

Now we are ready to define the evaluation set we will use to construct the code. Let $\infty_2$ denote the point at infinity of $E_2$. Consider the following collection of $\ell$ rational points on $E_2$ 
$$S = \{p_1, \ldots, p_\ell\} \subseteq E_2(\F_q) \setminus \{\infty_2\},$$
such that for every $i \in \{1, \ldots, \ell\}$ there exists an $\F_q$-rational point in the pre-image $\psi^{-1}(p_i)$. The evaluation set of our code will be 
\begin{equation}\label{eq:evaluationsetHLRC}
    T = \bigcup_{i=1}^\ell \psi^{-1}(p_i) \subseteq E(\F_q) \setminus E[m].
\end{equation}
The inclusion $T \subseteq E(\F_q)$ follows from two assumptions we have mentioned before: given a point $p_i \in S$, its preimage is the set $\psi^{-1}(p_i)=\{\overline{p_i} \oplus E[m]\} \subseteq E(\overline{\F_q})$. Since $E[m] \subseteq E(\F_q)$ and $\overline{p_i} \in E(\F_q)$ for every $i \in \{1, \ldots, \ell\}$, we have $\psi^{-1}(p_i) \subseteq E(\F_q)$ because if $p,q \in E(\F_q)$, then $p \oplus q \in E(\F_q)$.

Now we define the vector space $V$ of rational functions to be evaluated at the points of $T$: in this case, we construct a basis for $ V$. Consider a collection of rational functions
$$\langle h_1, \ldots, h_t \rangle = H^0(E_2, \OO_{E_2}(t \infty_2)),$$
such that $0 < t \leq \ell - m + 1$. In fact, we will see that the global minimum distance is affected by $t$ and our estimates for a lower bound (see \Cref{construction_curves: section 4}) will motivate the restrictions on the choice of this integer. We can assume, without loss of generality, that the functions $h_i$ have strictly distinct pole orders at $\infty_2$. Recalling that for a group $G < E(\F_q)$, we denote by $D_G$ the divisor defined as the formal sum of the points in $G$, let $f \in \F_q(E)$ and $g \in \F_q(E_1)$ be two rational functions satisfying the following conditions:
\begin{enumerate}
    \myitem{(f.$1$)}\label{1a)} $\F_q(E) = \F_q(E_1)(f)$;
    \myitem{(f.$2$)}\label{1b)} $f \in H^0(E, \OO_{E}(D_{G_1}))$;
    \myitem{(g.$1$)}\label{2a)} $\F_q(E_1) = \F_q(E_2)(g)$;
    \myitem{(g.$2$)}\label{2b)} $g \in H^0(E_1, \OO_{E_1}(D_{\overline{G_2}}))$.
\end{enumerate}

 Note that the conditions on $f$ and $g$ as functions contained in the aforementioned Riemann--Roch spaces, respectively associated to $D_{G_1} \in \DDiv_{\F_q}(E)$ and $D_{\overline{G_2}} \in \DDiv_{\F_q}(E_1)$, can be satisfied, as follows from \Cref{prop: primitive_element_extension_group}. Moreover, as a consequence of these conditions, we have that the poles of $f$ and $\varphi_1^\ast(g)$ are all contained in $E[m]$, which by definition avoids the evaluation set $T=\psi^{-1}(S)$. This fact allows us to evaluate these functions at the points of $T$, obtaining a value in $\F_q$.

However, the aforementioned conditions are not enough for the recovery procedure of our code: indeed, the interpolation process requires $f$ and $g$ to separate respectively the points of any fiber of $\varphi_1$ and the $m$ local $\varphi_1$-fibers contained in a $\psi$-fiber\footnote{Since $g \in \F_q(E)$ is in fact the function $\varphi_1^\ast(g)$, it is constant on every $\varphi_1$-fiber. We need $g$ to separate those fibers within a $\psi$-fiber: \ie if $a \neq b \in \psi^{-1}(p_i)$ are such that $\varphi_1(a) \neq \varphi_1(b)$, then we want $g(a) \neq g(b)$.}. In addition, we will require $f$ to be injective over a $\psi$-fiber, to maximize the minimum distance of the middle codes. Thus, the necessary further assumptions on $f$ and $g$ are the following:

\begin{enumerate}
    \myitem{(f.$3$)}\label{1c)} $f(p \oplus s) \neq f(p)$ for every $p \in T$ and $s \in E[m] \setminus \{\infty_E\}$,
    \myitem{(g.$3$)}\label{2c)} $\big(\varphi_1^\ast(g)\big)(p \oplus s_2) \neq \big(\varphi_1^\ast(g)\big)(p)$ for every $p \in T$ and $s_2 \in G_2 \setminus \{\infty_E\}$.
\end{enumerate}

Summarizing the above, we would like $f$ and $g$ to satisfy conditions \ref{1a)}, \ref{1b)}, \ref{1c)} and \ref{2a)}, \ref{2b)}, \ref{2c)}, respectively. 

\begin{rem}\label{rem: difference implies primitive element}
    In fact, once we obtain \ref{1c)}  and \ref{2c)} , the conditions \ref{1a)}  and \ref{2a)}  can be proved superabundant. As we have recalled in \Cref{prop: primitive_element_extension_group}, one can characterize the primitive element $f$ (or $g$) saying that it has trivial stabilizer in $G_1$ (or $\overline{G_2}$), \ie for every non-trivial $s_1 \in G_1$ (or $s_2 \in \overline{G_2})$ there exists a point $p \in E$ (or $p \in E_1)$ such that $f(p \oplus s_1) \neq f(p)$ (or $g(p \oplus s_2) \neq g(p))$. This immediately follows from condition \ref{1c)} (equivalently, \ref{2c)}). Henceforth, in what follows, we will only care about conditions \ref{1b)} and \ref{1c)}  (\ref{2b)} and \ref{2c)}, respectively). 
\end{rem}

    The existence of rational functions $f$ and $g$ satisfying the aforementioned assumptions is not always guaranteed. We must impose an additional condition on the size of the field $\F_q$, which we make explicit below. 

\begin{lem}\label{rem: existence f g}
    Suppose that the size of the field $\F_q$ is subject to the following inequality: 
    \begin{equation}\label{f_condition on q}
        q^m - 1 > \ell \cdot q^{m-1} \cdot \binom{m^2}{2}.
    \end{equation}
    Then the existence of rational functions $f$ and $g$ satisfying the aforementioned conditions is guaranteed.
\end{lem}

\begin{myproof} 
    Firstly, let us consider the conditions on $f$. From \ref{1b)}, we have that $f$ is contained in a vector space over $\F_q$ of dimension given by the degree of the divisor $G_1$, which is exactly $m$, thus, there are exactly $q^m-1$ different non-trivial rational functions in $H^0(E,\OO_E(D_{G_1}))$. The condition \ref{1c)} can be translated saying that for every $p_i \in S$ and for every $a \neq b \in \psi^{-1}(p_i)$, we must have $f(a) \neq f(b)$. For every $p_i \in S$ this gives exactly $\binom{m^2}{2}$ conditions, thus $f$ is contained in the complementary of the union $U$ of the hyperplanes given by the conditions $\bigg\{\{h(a)=h(b)\}_{a,b \in \psi^{-1}(p_i)}\bigg\}_{i=1}^\ell$. The subset $U \subseteq H^0(E,\OO_E(D_{G_1}))$ has at most cardinality $\ell \cdot q^{m-1} \cdot \binom{m^2}{2}$, since every hyperplane in $H^0(E,\OO_E(D_{G_1}))$ has cardinality $q^{m-1}$. Therefore, the existence of such a function is guaranteed if $\# \big(H^0(E,\OO_E(D_{G_1})) \setminus \{0\}\big) > \# U$, \ie if
    \begin{equation}\label{f_condition on q}
        q^m - 1 > \ell \cdot q^{m-1} \cdot \binom{m^2}{2}.
    \end{equation}
    With analogous arguments, in the case of $g \in H^0(E_1, \OO_{E_1}(D_{\overline{G_2}})),$ we just have $\binom{m}{2}$ conditions for every point $p_i \in S$, hence we just need
    \begin{equation}\label{g_condition on q }
        q^m - 1 > \ell \cdot q^{m-1} \cdot \binom{m}{2}.
    \end{equation}
    In conclusion, the existence of the rational functions $f$ and $g$ as above is guaranteed if the inequality in \Cref{f_condition on q} holds. 
\end{myproof}
\begin{rem}
We point out that one can restate the condition from the theorem above as follows:
    \begin{equation*}
        q^m \gg \ell \cdot q^{m-1} \cdot \binom{m^2}{2} \Longleftrightarrow q \gg \ell \cdot \binom{m^2}{2} = \ell \cdot \frac{m^2 \cdot (m^2-1)}{2} \sim \ell \cdot \frac{m^4}{2}.
    \end{equation*}
    \end{rem}

We are finally ready for the explicit construction of the code on our quotient of elliptic curves.

\begin{prop}\label{construction_curves: section 4}

    Consider an integer $m > 3$ and an elliptic curve $E$ defined over a field $\F_q$ such that $q \gg \ell \cdot \frac{m^2 (m^2-1)}{2}$, $\gcd(m,q)=1$ and $E[m] \subseteq E(\F_q)$. We fix the subgroups $G_1, G_2 < E[m]$ and the corresponding isogenies $\varphi_1$ and $\varphi_2$, the evaluation set $T \subseteq E(\F_q)$ and the functions $f, g, h_i$ satisfying all the assumptions presented in the previous paragraphs. Let $r$ be defined as $r=m-1$ and let $V$ be the following $\F_q$-vector space of rational functions:
    $$V = \langle h_i \cdot g^j \cdot f^h \ | \ 1 \leq i \leq t, \ 0 \leq j \leq r-1, \ 0 \leq h \leq r-1 \rangle\subseteq\F_q(E).\footnote{Note that we simply denote by $h_i$ and $g$ the rational functions $\psi^\ast(h_i)$ and $\varphi_1^\ast(g)$, which are contained in $\F_q(E)$ since we have the tower of function field extensions
    $\F_q(E_2) \subseteq \F_q(E_1) \subseteq \F_q(E).$}$$
    The corresponding AG code $\mathcal{C}(E,\{\varphi_1, \varphi_2\}, V,T)$ is the image of the evaluation map
    \[
        \funcDef{\ev_T}{V}{\F_q^n}
        {\delta}{(\delta(p))_{p \in T}}
    \]
    where $n = \#T = \ell \cdot (r+1)^2 = \ell \cdot m^2$. This is an HLRC, whose middle and lower layers can be presented as follows:

\begin{itemize}
    \item for every point $p \in T$ there exists $p_i \in S$ such that $\psi(p) = p_i$. The middle code supported on $p$ has evaluation set given by $\psi^{-1}(p_i) = p \oplus E[m]$. On this fiber, the vector space becomes
    $$V_1 = \langle g^j \cdot f^h \ | \ 0 \leq j \leq r-1, \ 0 \leq h \leq r-1 \rangle.$$
    
    \item Given a point $p \in T$, the corresponding lower code has evaluation set given by $\varphi_1^{-1}(\varphi_1(p)) = p \oplus G_1$. If we see the middle code described above as an LRC, this lower code is supported on the recovery set of the point $p$ and the vector space becomes
    $$V_2 = \langle f^h \ | \ 0 \leq h \leq r-1 \rangle.$$
\end{itemize}

The first two parameters of the middle and lower codes, \ie length and dimension, are respectively:
\begin{itemize}
    \item $n_1 = m^2 = (r+1)^2$ and $k_1 = r^2$,
    \item $n_2 = m = r+1$ and $k_2 = r$.
\end{itemize}

Moreover, the following estimates for the minimum distances hold:
\begin{itemize}
    \item $d_1 \geq n_1 - \deg_\psi(g) \cdot (r-1) - \deg_\psi(f) \cdot (r-1)$, where $\deg_\psi(f)$ and $\deg_\psi(g)$ are the maximum number of times the functions $f, g: E \longrightarrow \p^1$ assume any value in $\F_q$ on a fiber of $\psi$.
    \item $d_2=2$.
\end{itemize}

Finally, the full code has the following parameters:
\begin{itemize}
    \item $n=\ell \cdot m^2$;
    \item $k= t \cdot r^2$;
    \item $d \geq n - \big(\max_i\{\deg_E(h_i)\} + \deg_E(g) \cdot (r-1) + \deg_E(f) \cdot (r-1)\big)$.
\end{itemize}
\end{prop}

\begin{myproof}
    The proof of this proposition consists, in fact, of showing that the estimates for the parameters are correct. According to the definition of the code, the global length of the code is $n=\ell \cdot m^2$ and $n_1 = m^2$, $n_2 = m$. Furthermore, since the evaluation set of a middle code is a fiber of the isogeny $\psi$ and the functions $h_i$ are pull-backs of the elements of a basis for $H^0(E_2, \OO_{E_2}(t \cdot \infty_2))$, they are constant on a $\psi$-fiber. Hence, the restriction of $V$ on the evaluation set of a middle code is $V_1$. One can apply the same argument for the vector space $V_2$ defined over the evaluation set of the lower codes. As a consequence, if we show that the evaluation maps corresponding to the middle and lower codes are injective, the values of $k$, $k_1$ and $k_2$ are straightforward to prove. We obtain the injectivity of those maps by presenting strictly positive lower bounds for the minimum distances $d$, $d_1$ and $d_2$. 

    As stated in \cite[Prop.~IV.1]{ballentine2019}, every lower code is a single parity check code, with length $r+1$ and dimension $r$, hence the minimum distance is $d_2=2$. The minimum distance of the middle code is bounded from below by the difference between its length and the maximum number of zeros of a function in $V_1$ on a fiber of $\psi$. This second number equals to the sum of the contributions given by $f$ and $g$ on every fiber of $\psi$. Then, we obtain exactly  $\deg_\psi(g) \cdot (r-1) + \deg_\psi(f) \cdot (r-1)$: indeed, a non-trivial function $\delta \in V_1$ can be expressed as:
    $$\delta(f,g) = A_0(g) + A_1(g) f + \ldots + A_{r-1}(g)f^{r-1},$$
    where $A_i(g)$ are polynomials in $g$ of maximum degree $r-1$. Now, we can count the maximum number of local fibers over which $\delta$ is identically zero. Later, on the remaining fibers, we give an estimate for the maximum number of individual zeros of the function. Since $g$, by definition and condition \ref{2c)}, is constant on every $\varphi_1$-fiber and separates them, for the function $\delta$ to vanish on a local fiber, we require all the coefficients $A_i(g)$ to be simultaneously zero. The degree of $A_i(g)$ as a polynomial in $g$ is at most $r-1$, hence these coefficients share at most $r-1$ common roots. As a consequence, the function $\delta$ can vanish on at most $r-1$ entire $\varphi_1$-fibers, contributing to $\deg_\psi(g) \cdot (r-1)$ zeros. On each of the remaining $2$ local fibers, the function $\delta$ can be seen as a non-trivial polynomial in $f$ of maximum degree $r-1$. By condition \ref{1c)}, the function $f$ is injective on a $\psi$-fiber, \ie $\deg_\psi(f)=1$, thus $\delta$ has at most $r-1$ isolated zeros distributed across the remaining $\varphi_1$-fibers, which can be expressed as $\deg_\psi(f) \cdot (r-1)$, since $\deg_\psi(f)=1$. Hence, since $n_1 > \big(\deg_\psi(g) + \deg_\psi(f)\big)\cdot (r-1)$, the evaluation map corresponding to a middle code is injective, and the final estimate for the minimum distance is
    $$d_1 \geq m^2 - m \cdot (m-2) - (m-2) = m + 2.$$
    For the global code, we can use the same arguments we have expressed above for middle codes, but our estimates for the maximum number of zeros depend also on the functions $h_i$, and the degrees of the functions $f, g: E \longrightarrow \p^1$ replace their degrees on a fiber of $\psi$. Therefore, using the correct values we obtain
    \begin{align*}
        d &\geq \ell \cdot m^2 - \big(t \cdot m^2 + m^2 \cdot (r-1) + m \cdot (r-1) \big) \\
        &= m^2 \cdot (\ell - t - m + 2) - m \cdot (m-2) \\
        &= m^2 \cdot (\ell - t - m +1) + 2m \geq 2m,
    \end{align*}
    where the last inequality follows from the fact that we have chosen $t \leq \ell - m +1$.
    
\end{myproof}
\subsection{HLRCs on elliptic surfaces and their parameters}\label{construction_surfaces: section 4}

Consider a projective curve $B$ of genus $g_B$ defined over $\F_q$ and an elliptic $\F_q$-surface $\pi: \E \longrightarrow B$. As we have seen in \Cref{prop_corr_ell_curves_surfaces}, there exists a corresponding elliptic curve $E$ defined over the field $K=\F_q(B)$, which is in fact the generic fiber of $\E$ over $B$, for which we have an isomorphism $\F_q(\E) \cong K(E)$.

We are going to repeat the construction exposed in \Cref{subsection: hlrc quotients elliptic curves} on the generic fiber of $\E$: as a consequence, we will obtain a code on the corresponding elliptic surface such that its restriction to suitable elliptic fibers gives us exactly what we achieved in the previous subsection. In fact, every point of the generic fiber is a section of the elliptic fibration, which can be thought of as a family of points indexed by the elements of $B(\F_q)$. 

\smallskip
Consider an integer $m > 3$ coprime with $q$ and suppose that $E[m] \subseteq E(K)$: geometrically, on the surface, this means that the $m$-torsion points of the elliptic fibers form an $\F_q$-rational section on $\E$.

Take two independent points $m_1$ and $m_2$ in $E(K)[m]$ of order $m$ and the corresponding subgroups $G_1 = \langle m_1 \rangle$ and $G_2= \langle m_2 \rangle$. Thus, we can define the following sequence of separable $K$-isogenies:
$$E \stackrel{\varphi_1} \longrightarrow E_1=E/G_1 \stackrel{\varphi_2}\longrightarrow E_2=E_1/\overline{G_2},$$
with $\psi= \varphi_2 \circ \varphi_1$ and $\overline{G_2}=\varphi_1(G_2) < E_1(K)$.

Consider a collection of points 
$$\tilde{S} = \{p_1, \ldots, p_\ell\} \subseteq E_2(K) \setminus \{\infty_2\},$$
such that for every $i \in \{1, \ldots, \ell\}$ there exists a $K$-rational point in the pre-image $\psi^{-1}(p_i)$. The evaluation locus\footnote{The set $\tilde{T}$ is a collection of points in the generic fiber $E$, which corresponds to a collection of sections in the elliptic surface $\E$. The evaluation set of the code will be the union of the supports of the sections in $\tilde{T}$.} of the code is
$$\tilde{T} = \bigcup_{i=1}^\ell \psi^{-1}(p_i) \subseteq E(K) \setminus E[m].$$
The next goal is to define a vector space of appropriate rational functions contained in $K(E) = \F_q(\E)$. As we have seen in \Cref{subsection: hlrc quotients elliptic curves}, we want two functions $f \in K(E)$ and $g \in K(E_1)$ such that:
\begin{enumerate}

    \myitem{(f.$2$)}\label{1abis} $f \in H^0(E, \OO_{E}(D_{G_1}))$;
    \myitem{(f.$3$)}\label{1bbis} $f(p \oplus s) \neq f(p)$ for every $p \in \tilde{T}$ and $s \in E[m] \setminus \{\infty_E\}$;
    \myitem{(g.$2$)}\label{2abis} $g \in H^0(E_1, \OO_{E_1}(D_{\overline{G_2}}))$;
    \myitem{(g.$3$)}\label{2bbis} $\big(\varphi_1^\ast(g)\big)(p \oplus s_2) \neq \big(\varphi_1^\ast(g)\big)(p)$ for every $p \in \tilde{T}$ and $s_2 \in G_2 \setminus \{\infty_E\}$.
\end{enumerate}

Recall that the conditions \ref{1bbis} and \ref{2bbis} imply that the functions $f$ and $g$ are primitive elements for the respective field extensions. The proof for the existence of these two functions is substantially the same as we presented in  \Cref{rem: existence f g}, but there is no need for additional hypotheses on the size of $q$. Indeed, the Riemann--Roch spaces $H^0(\OO_{E}(D_{G_1}))$ and $H^0(E_1, \OO_{E_1}(D_{\overline{G_2}}))$ are vector space over the infinite field $K=\F_q(B)$. A finite union of proper vector subspaces cannot cover a $K$-vector space. Thus, the existence of $f$ and $g$ as above is guaranteed without any further assumption on the size of $q$.

There is another difference between the aforementioned conditions \ref{1bbis} and \ref{2bbis} and the analogous conditions in \Cref{subsection: hlrc quotients elliptic curves}: in this case, in fact, the evaluation of a rational function in $K(E)$ at a point $p \in E(K)$ is an element of $K$, \ie a rational function defined over the algebraic curve $B$. Thus, these two conditions can be restated as:
    \begin{enumerate}
    \item[(f.$3$)] $\delta_{f,p,s}=f(p \oplus s) - f(p) \neq 0 \in \F_q(B)$ for every $p \in \tilde{T}$ and $s \in E[m] \setminus \{\infty_E\}$,
    \item[(g.$3$)] $\delta_{g,p,s_2}=\big(\varphi_1^\ast(g)\big)(p \oplus s_2) - \big(\varphi_1^\ast(g)\big)(p) \neq 0 \in \F_q(B)$ for every $p \in \tilde{T}$ and $s_2 \in G_2 \setminus \{\infty_E\}$.
    \end{enumerate}
We shall use these $\delta_{f,p,s}$ and $\delta_{g,p,s}$ in what follows.
\smallskip

Now, take $\langle h_1, \ldots, h_t \rangle = H^0(E_2, \OO_{E_2}(t \infty_2))$ such that $0 < t \leq \ell - m +1$. In addition, we can choose these functions to have strictly distinct pole orders at $\infty_2$. We are ready to define an appropriate evaluation set using the collection of sections $\tilde{T}$.

Firstly, we want a set $C \subseteq B(\F_q)$ of points of good reduction for the elliptic fibration: this is to make sure that if we restrict the global code on the surface $\E$ to one of the fibers $E_\gamma$, with $\gamma \in C$, we will obtain the code built in  \Cref{subsection: hlrc quotients elliptic curves}. For this reason we want to avoid the elements $\gamma \in B(\F_q)$ such that:
\begin{enumerate}
    \item $E_\gamma$ is not an elliptic curve,
    \item $\gamma$ is a pole for one of the coefficients in $K$ defining $f, g$ or $h_i$ as elements of the respective Riemann--Roch spaces,
    \item there exists a section $p \in \tilde{T}$ and an element $s \in E[m] \setminus \{\infty_E\}$ such that $\delta_{f,p,s}(\gamma)=0$ or $\delta_{g,p,s}(\gamma)=0$.
\end{enumerate}

Consider $B_{\text{bad}} \subset B(\F_q)$ as the set of points satisfying one of the above three conditions, and let us define $C= B(\F_q) \setminus B_{\text{bad}}$ (see \Cref{rem:bad_points} for a discussion on the size of this set $C$). The evaluation set of the code over the elliptic surface $\E$ is 
$$T = \bigcup_{\gamma \in C} T_\gamma,$$
with $T_\gamma = \{p_\gamma=\text{sp}_\gamma(p) \in E_\gamma(\F_q) \ | \ p \in \tilde{T}\}$, where $\Sp_\gamma: E(K) \longrightarrow E_\gamma(\F_q)$ is the specialization map. 

Before presenting the final result of this construction, we recall the definition of LRC with $H \geq 2$ hierarchical levels, as stated in \cite[\S~4, Def.~6]{Malmskog2025}. In fact, this is a generalization of Definition \ref{def hierarchical LRC codes}.

\begin{defn}\label{def: HLRC multiple}
    Let $H \geq 2$, $n_j, k_j, d_j$ be positive integers for every $j= 1, \ldots, H$, such that:
    \begin{itemize}
        \item $n_1 > \ldots > n_H$,
        \item $k_1 \geq \ldots \geq k_H$,
        \item $d_1 > \ldots > d_H$.
    \end{itemize}
    An $[n,k,d]$-linear code $\mathcal{C}$ has \textit{$H$-level hierarchical locality with parameters}
    $$[(n_1, k_1, d_1), \ldots, (n_H, k_H, d_H)]$$
    if for every $i \in \{1, \ldots, n\}$ there exists a collection of $H$ punctured codes $\{(\C_i)_j\}$ with $j \in \{1, \ldots, H\}$, each of them of length $n_j$, whose supports contain the coordinate $i$. Moreover, if we define $I_{j,i} = \Supp((\C_i)_j) \setminus \{i\}$, we have that:
    \begin{itemize}
        \item $I_{H,i} \subseteq \ldots \subseteq I_{1,i}$;
        \item $\dim((\C_i)_j) \leq k_j$ for every $j$;
        \item the minimum distance of $(\C_i)_j$ is at least $d_j$ for every $j$;
        \item $(\C_i)_j$ has $(H-j)$-level hierarchical locality through the collection of codes $\{(\C_i)_j \ : \ j < k \leq H\}$.
    \end{itemize}
\end{defn}

\begin{rem}
    In the definition above, one can choose $\C_1$ to be $\C$, but this is not necessary: we want $\C_1$ to correspond to the first effective layer for the local recovery. In the easier case, that is, $H=2$, the collection $\{(\C_i)_1\}$ corresponds to the set of middle codes associated with the coordinate $i$, while $\{(\C_i)_2\}$ is the set of lower codes.
\end{rem}
\begin{thm}\label{prop: parameters_surfaces section 4}
    Consider a projective $\F_q$-curve $B$ of genus $g_B$ and an elliptic $\F_q$-surface $\pi: \E \longrightarrow B$, an integer $m > 3$ coprime with $q$ and two collections of points $T \subseteq \E(\F_q)$ and $C \subseteq B(\F_q)$ defined as above. Let $D \in \DDiv_{\F_q}(B)$ be an effective divisor of degree $\delta$, with $g_B \leq \delta < \#C$, such that $C \cap \Supp(D) = \emptyset$. Let $V \subseteq \F_q(\E)$ be the following $\F_q$-vector space:
    $$V = \bigg\{\sum_{i,j,h} b_{i,j,h} \cdot h_i \cdot g^j \cdot f^h \ | \ b_{i,j,h} \in H^0(B,\OO_B(D)), \ 1 \leq i \leq t, \ 0 \leq j, h \leq r-1\bigg\},$$
    where $f, g, h_i$ are chosen as above. 
    The corresponding AG code $\C(\E,V,T)$ has $3$-level hierarchical locality. The first layer of the code is supported on the elliptic fibers $E_\gamma$, with $\gamma \in C$, and the corresponding codes are exactly as the ones we have built in \Cref{subsection: hlrc quotients elliptic curves}. The parameters of the global code are:
    \begin{itemize}
        \item $n =\# T= \# C \cdot \ell \cdot m^2$,
        \item $k = h^0(B,D) \cdot t \cdot r^2$,
        \item $d \geq (\#C - \delta) \cdot \big(m^2 \cdot (\ell - t - m + 1) + 2m \big)$.
    \end{itemize}
\end{thm}
\begin{myproof}
    We are going to prove that, for any fixed $\gamma \in C$, the punctured code $\C(E_\gamma, V_\gamma, T_\gamma)$ is exactly the one built in \Cref{subsection: hlrc quotients elliptic curves} on an elliptic curve. It will then follow that $\C(\E,V,T)$ meets \Cref{def: HLRC multiple} with $H=3$. Moreover, the length and the dimension of the global code are straightforward to prove, while the estimate for the minimum distance is performed as in \Cref{prop: parameters_LRC(2)_ell_surfaces}, since $m^2 \cdot (\ell - t-m+1) + 2m$ is a lower bound for the minimum distance on an elliptic curve, as seen at the end of \Cref{construction_curves: section 4}.
    
   For a fixed $\gamma \in C$, we divide the proof in two parts:
    \begin{enumerate}
        \item $T_\gamma=\text{sp}_\gamma(\tilde{T})$ is exactly the evaluation set of the code in \Cref{subsection: hlrc quotients elliptic curves},
        \item the restrictions $f_\gamma, g_\gamma$ and $h_{i,\gamma}$ of the functions $f, g$ and $h_i$ in $K(E)$ satisfy respectively the conditions \ref{1b)} and \ref{1c)},  \ref{2b)} and \ref{2c)} and analogous conditions for the $h_i$.
    \end{enumerate}

   The first claim is a consequence of  \Cref{thm: specialization map}. Indeed, the specialization map is a group homomorphism, thus, the image of $G_1, G_2 < E(K)[m]$ are the independent subgroups $G_{1,\gamma}, G_{2,\gamma}$ which generate the $m$-torsion group on $E_\gamma(\F_q)$. Hence, $T_\gamma$ is a union of $E_\gamma[m]$-cosets, as in \Cref{subsection: hlrc quotients elliptic curves}. Moreover, $\Sp_\gamma$ is injective on $E(K)[m]$, thus if $p = q \oplus s \in \tilde{T}$, with $\infty_E \neq s \in E(K)[m]$, then
   $p_\gamma = q_\gamma \oplus s_\gamma \neq q_\gamma$ since $s_\gamma =\Sp_\gamma(s) \neq \Sp_\gamma(\infty_E) = \infty_\gamma$. This proves that the specialization map is also a bijection on $\tilde{T}$, and $T_\gamma$ is the evaluation set we wanted on the elliptic curve $E_\gamma$, with $\#T_\gamma = \ell \cdot m^2$.

   The second claim is a consequence of the conditions imposed on the functions $f, g, h_i \in K(E)$ and on the set $C \subseteq B(\F_q)$. The proof for the function $f_\gamma$ is similar to the one for $g_\gamma$, hence we will work only on $f_\gamma$. Later, we will treat the case of $h_{i,\gamma}$. 
   
   We want to show that $f_\gamma \in H^0(E_\gamma, \OO_{E_\gamma}(D_{G_{1,\gamma}}))$ and it is injective on a $E_\gamma[m]$-coset contained in $T_\gamma$. Since $\gamma \in C$, the function $f$ cannot have the fiber $E_\gamma$ as a pole and we have that\footnote{The equality between those divisors of poles follows immediately from the definition of Cartier divisors, given for example in \cite[Ch.~II, \S 6]{Hartshorne}, and the definition of the restriction of a Cartier divisor to a closed subscheme, which is not entirely contained in the support of the divisor: indeed, in our case $E_\gamma \subset \E$ is a closed subscheme which satisfies the hypothesis above.} 
   $$\ddiv_\infty(f_\gamma) = \ddiv_\infty(f)_{|E_\gamma}.$$
   As a consequence, $f \in H^0(E, \OO_E(D_{G_1})) $ implies $ f_\gamma \in H^0(E_\gamma, \OO_{E_\gamma}(D_{G_{1,\gamma}})).$
   
   In addition, on the generic fiber we have imposed $\delta_{f,p,s} \in \F_q(B)$ to be non-identically zero for every $p \in \tilde{T}$ and $s \in E(K)[m]$. Since $\gamma \in C$, $\gamma$ is not a zero of $\delta_{f,p,s}$ and 
   $$0 \neq \delta_{f,p,s}(\gamma) = f(p \oplus s)(\gamma) - f(p)(\gamma) =f_\gamma(p_\gamma \oplus_\gamma s_\gamma) - f_\gamma(p_\gamma),$$
   where the last equality follows from \Cref{rem: equivalence evaluation}. This proves the injectivity of $f_\gamma$ on a coset contained in $T_\gamma$.
   
   The restricted functions $\{h_{1,\gamma}, \ldots, h_{t,\gamma}\}$ form a basis of the $\F_q$-vector space $H^0(E_{\gamma,2},\OO_{E_{\gamma,2}}(t \cdot \infty_{\gamma,2}))$. Indeed, since $\gamma \in C$, we have that the restriction of the function $h_i$ to the elliptic fiber $E_\gamma$ preserves the multiplicity of $\infty_{\gamma,2}$ as a pole of $h_{i, \gamma}$ and having distinct pole orders implies linear independence. This concludes the proof of the proposition. 
\end{myproof}

\begin{rem}\label{rem:bad_points}
The existence of a non-empty subset $C \subseteq B(\mathbb{F}_q)$ is a crucial requirement for our construction. Since we have defined it as the complement of $B_{\mathrm{bad}}$ in $B(\F_q)$, we want to estimate the maximum number of points that we must exclude. The set $B_{\mathrm{bad}}$ consists of three types of points:
\begin{enumerate}
    \item points $\gamma$ such that the fiber $E_\gamma$ is singular. If the elliptic surface $\mathcal{E}$ is defined by a Weierstrass equation with coefficients $a_i \in \mathbb{F}_q(B)$, these points correspond to the roots of the discriminant $\Delta(a_i)$, whose number is bounded by $\deg(\Delta)$;
    \item points $\gamma$ that are poles of the coefficients in $K=\F_q(B)$ defining $f_i, g_i$, or $h_j$. Since these functions belong to specific Riemann--Roch spaces, the number of their poles is bounded by the degrees of their associated divisors;
    \item points $\gamma$ where the separating functions vanish, \ie $\delta_{f_i,p,s}(\gamma) = 0$ or $\delta_{g_i,p,s}(\gamma) = 0$. These are non-identically zero rational functions in $\mathbb{F}_q(B)$, thus each of them can vanish on at most a finite number of points, bounded by their degree.
\end{enumerate}

As a consequence, we have a finite upper bound $M$ for the cardinality of $B_{\mathrm{bad}}$. It is crucial to note that $M$ is an absolute geometric constant that strictly depends on the invariants of the surface, such as, for example, the degree of the discriminant and the chosen degrees of the divisors, which are independent of the field size $q$. 

By the Hasse--Weil bound, the number of rational points on the base curve $B$ satisfies $\#B(\mathbb{F}_q) \ge q + 1 - 2g_B \sqrt{q}$. Therefore, to guarantee that $C = B(\mathbb{F}_q) \setminus B_{bad}$ contains at least $g_B + 1$, as required in the statement of \Cref{prop: parameters_surfaces section 4}, we need
\begin{equation*}
    q + 1 - 2g_B \sqrt{q} > M + g_B.
\end{equation*}

Since $M$ and $g_B$ are fixed constants, one can always satisfy this inequality by passing to a sufficiently large constant extension of the base field $\mathbb{F}_{q}$. Over this field extension, the upper bound $M$ remains constant, while the number of rational points grows asymptotically as $q^k$. Furthermore, we can maximize the size of $C$ (and consequently the length of the global code), choosing $B$ to be a maximal curve, \ie a curve which attains the Hasse--Weil upper bound $$\#B(\mathbb{F}_{q}) = q + 1 + 2g_B \sqrt{q},$$ thus providing the maximum possible number of $\F_q$-rational points.
\end{rem}

\section{Combining availability and hierarchical locality}\label{section: HLRC with availability}
The aim of this last section is to achieve availability and hierarchical locality together on elliptic curves and surfaces. To start with, we recall the definition of a generalized version of an LRC with $H$-level hierarchical locality and availability, as presented in \cite[\S 5]{Malmskog2025}. 

\begin{defn}\label{def: H-level hierarchy with availability}
	Let $H, n_j, t_j, k_j, d_j, r_{j,a}, \rho_{j,a}$ be positive integers for every $j \in \{1, \ldots, H\}$ and $a \in \{1, \ldots, t_j\}$, such that:
	\begin{itemize}
		\item $H \geq 2$;
		\item $k_1 \geq \ldots \geq k_H$;
		\item $d_1 > \ldots > d_H$. 
	\end{itemize}
	Let $\C$ be a $[n,k,d]$-linear code with $H$-level hierarchical locality with parameters $(n_j,k_j,d_j)$ and, for every $i \in \{1,\ldots, n\}$, let $(\C_i)_j$ be the $j$-th level middle code for the $i$-th coordinate. If for every $i$, $(\C_i)_j$ is an $(r_{j,a}, \rho_{j,a})$-LRC with availability $t_j$, the code $\C$ is said to have $H$-level hierarchy with availability $(t_j)$. 
\end{defn}

\subsection{Hierarchical locality and availability on elliptic curves}\label{construction_curves: section 5}

In this subsection, we will present the crucial construction on elliptic curves for achieving hierarchical locality and availability on elliptic surfaces, similarly to what we have done in the previous section.

\smallskip
Consider an elliptic curve $E$ defined over $\F_q$ and an integer $m = v \cdot w > 0$ such that $\gcd(v,w)=1$, $m$ is coprime with $q$, $q \geq m$ and $E[m] = E[v] \times E[w] \subseteq E(\F_q)$. The $m$-torsion subgroup is isomorphic to 
$$\Z/m \times \Z/m \cong (\Z/v \times \Z/w) \times (\Z/v \times \Z/w) \cong (\Z/v)^2 \times (\Z/w)^2.$$
Again, we have mentioned our initial assumptions on the size of $q$, which will then be subject to other conditions in order to guarantee a well-posed definition for our code.
Take $p_{v,1}$ and $p_{v,2}$, two independent points of order $v$, and $p_{w,1}$ and $p_{w,2}$, two independent points of order $w$. We will denote by $G_{v,1}$, $G_{v,2}$, $G_{w,1}$ and $G_{w,2}$ the subgroups of $E[m]$ generated by those points.
Let us consider the following elliptic curves:
\begin{itemize}
    \item $E_{v,i} = E/G_{v,i}$, $E_v = E/E[v]$,
    \item $E_{w,i} = E_v/\psi_v(G_{w,i})$, $E_m = E_v/\psi_v(E[w]) = E/E[m]$, where $\psi_v: E \longrightarrow E/E[v]$ is the isogeny corresponding to $E[v]$.
\end{itemize}
Note that the elliptic curve $E_{w,i}$, with $i=1,2$, is in fact the quotient $E/(E[v] \times G_{w,i})$. In the following paragraphs, we will denote $E[v] \times G_{w,i}$ by $H_{w,i}$. Moreover, we denote by $\psi_m$ the composition $\psi_w \circ \psi_v: E \longrightarrow E_m$.

The following diagram illustrates the situation:
% https://q.uiver.app/#q=WzAsNyxbMiwwLCJFIl0sWzAsMSwiRV97diwxfSJdLFs0LDEsIkVfe3YsMn0iXSxbMiwyLCJFX3YiXSxbMCwzLCJFX3t3LDF9Il0sWzQsMywiRV97dywyfSJdLFsyLDQsIkVfbSJdLFswLDEsIlxcdmFycGhpX3t2LDF9IiwxXSxbMCwyLCJcXHZhcnBoaV97diwyfSIsMV0sWzAsMywiXFxwc2lfdiIsMV0sWzEsMywiXFxwc2lfe3YsMX0iLDFdLFsyLDMsIlxccHNpX3t2LDJ9IiwxXSxbMyw1LCJcXHZhcnBoaV97dywyfSIsMV0sWzMsNCwiXFx2YXJwaGlfe3csMX0iLDFdLFszLDYsIlxccHNpX3ciLDFdLFs0LDYsIlxccHNpX3t3LDF9IiwxXSxbNSw2LCJcXHBzaV97dywyfSIsMV1d
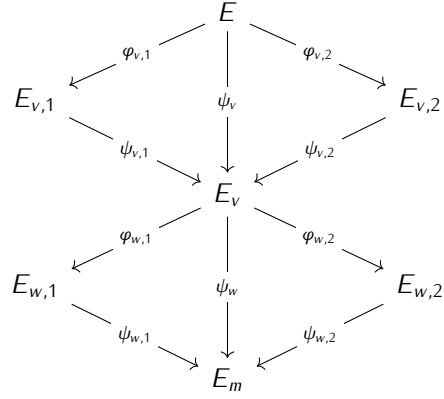
\[\begin{tikzcd}
	&& E \\
	{E_{v,1}} &&&& {E_{v,2}} \\
	&& {E_v} \\
	{E_{w,1}} &&&& {E_{w,2}} \\
	&& {E_m}
	\arrow["{\varphi_{v,1}}"{description}, from=1-3, to=2-1]
	\arrow["{\varphi_{v,2}}"{description}, from=1-3, to=2-5]
	\arrow["{\psi_v}"{description}, from=1-3, to=3-3]
	\arrow["{\psi_{v,1}}"{description}, from=2-1, to=3-3]
	\arrow["{\psi_{v,2}}"{description}, from=2-5, to=3-3]
	\arrow["{\varphi_{w,1}}"{description}, from=3-3, to=4-1]
	\arrow["{\varphi_{w,2}}"{description}, from=3-3, to=4-5]
	\arrow["{\psi_w}"{description}, from=3-3, to=5-3]
	\arrow["{\psi_{w,1}}"{description}, from=4-1, to=5-3]
	\arrow["{\psi_{w,2}}"{description}, from=4-5, to=5-3]
\end{tikzcd}\]

Consider a collection of points 
$$S = \{p_1, \ldots, p_\ell\} \subseteq E_m(\F_q) \setminus \{\infty_m\},$$
such that the pre-image of every point $\psi_m^{-1}(p_i)$ contains an $\F_q$-rational point. The evaluation set of the code is
$$T= \psi_m^{-1}(S) \subseteq E(\F_q) \setminus E[m].$$

Now take 
$$\langle h_1, \ldots, h_t \rangle = H^0(E_m, t \cdot \infty_m) \subseteq \F_q(E_m),$$
with $0 < t \leq \ell$ such that the functions $h_j$ have strictly distinct pole orders at $\infty_m$. Let $f_1, f_2, g_1, g_2$ be defined as follows:
\begin{itemize}
    \item for $i=1,2$, $\F_q(E_{v,i}) = \F_q(E_v)(f_i)$ such that:
    \begin{itemize}
        \myitem{($f_1.2$)} $f_1 \in H^0(E_{v,1}, \OO_{E_{v,1}}(D_{\overline{G_{v,2}}}))$;
        \myitem{($f_1.3$)} $\delta_{f_1,p,s_{v,2}} = \big(\varphi_{v,1}^\ast(f_1)\big)(p \oplus s_{v,2}) -  \big(\varphi_{v,1}^\ast(f_1)\big)(p) \neq 0 \in \F_q$ for every $p \in T$ and $s_{v,2} \in G_{v,2} \setminus \{\infty_E\}$;
        \myitem{($f_2.2$)} $f_2 \in H^0(E_{v,2}, \OO_{E_{v,2}}(D_{\overline{G_{v,1}}}))$,
      \myitem{($f_2.3$)}\label{cf23}$\delta_{f_2,p,s_{v,1}} = \big(\varphi_{v,2}^\ast(f_2)\big)(p \oplus s_{v,1}) -  \big(\varphi_{v,2}^\ast(f_2)\big)(p) \neq 0 \in \F_q$ for every $p \in T$ and $s_{v,1} \in G_{v,1} \setminus \{\infty_E\}$.
    \end{itemize}
    Here $\overline{G_{v,i}}$ is the image of the group $G_{v,i}$ in $E_{v,1}$ or $E_{v,2}$;
    \item for $i=1,2$, $\F_q(E_{w,i})=\F_q(E_m)(g_i)$ such that:
    \begin{itemize}
        \myitem{($g_1.2$)} $g_1 \in H^0(E_{w,1}, \OO_{E_{w,1}}(D_{\overline{G_{w,2}}}))$;
        \myitem{($g_1.3$)} $\delta_{g_1,p,s_{w,2}} = \big((\varphi_{w,1} \circ \psi_{v})^\ast(g_1)\big)(p \oplus s_{w,2}) -   \big((\varphi_{w,1} \circ \psi_{v})^\ast(g_1)\big)(p) \neq 0 \in \F_q$ for every $p \in T$ and $s_{w,2} \in G_{w,2} \setminus \{\infty_E\}$;
        \myitem{($g_2.2$)} $g_2 \in H^0(E_{w,2}, \OO_{E_{w,2}}(D_{\overline{G_{w,1}}}))$,
        \myitem{($g_2.3$)}\label{cg23} $\delta_{g_2,p,s_{w,1}} =  \big((\varphi_{w,2} \circ \psi_{v})^\ast(g_2)\big)(p \oplus s_{w,1}) -   \big((\varphi_{w,2} \circ \psi_{v})^\ast(g_2)\big)(p) \neq 0 \in \F_q$ for every $p \in T$ and $s_{w,1} \in G_{w,1} \setminus \{\infty_E\}$;
    \end{itemize} 
    Here $\overline{G_{w,i}}$, for $i=1,2$, is the image of the group $G_{w,i}$ in $E_{w,1}$ or $E_{w,2}$, respectively.
\end{itemize}

As always, we will simply write $f_i, g_i$ and $h_j$ instead of their respective pullbacks in $\F_q(E)$, where the tower of field extensions is the one corresponding to the diagram above. By definition, we know that:
\begin{itemize}
    \item $\deg(\varphi_{v,i})=\deg(\psi_{v,i})=v$, $\deg(\varphi_{w,i})=\deg(\psi_{w,i})=w$ for $i=1,2$,
    \item $\deg(\psi_v)=v^2$, $\deg(\psi_w)=w^2$ and $\deg(\psi_m)=m^2$.
\end{itemize}

\begin{rem}
	As we have seen in \Cref{rem: difference implies primitive element}, the conditions $(.3)$ imply that the functions $f_i$ and $g_i$, for $i=1,2$, are primitive elements for the respective field extensions. Moreover, one can apply the same arguments presented in \Cref{rem: existence f g} for the existence of these functions: in fact, we only need 
	$$q \gg \ell \cdot \max \bigg\{\binom{v}{2}, \binom{w}{2}\bigg\}.$$
\end{rem}
Let $V$ be the $\F_q$-vector space of rational functions on $E$ defined as
$$V = \langle h_i \cdot f_1^{a_1} \cdot f_2^{a_2} \cdot g_1^{b_1} \cdot g_2^{b_2}, \ | \ i=1, \ldots, t; \ a_j=0, \ldots, v-2; \ b_j = 0, \ldots, w-2\rangle.$$

We can state and prove the final proposition of this subsection, containing the results of the construction, together with the parameters of the code.
\begin{prop}\label{prop: parameters_curve section 5}
    The linear code $\C(E,V,T)$ obtained evaluating the functions of $V$ on the points of the evaluation set $T$ has a $2$-level hierarchy with availability $(2,2)$ according to \Cref{def: H-level hierarchy with availability}: every point $p_0 \in T$ is contained in the evaluation set of two distinct middle codes, supported on the non-trivial cosets of $H_{w,1}$ and $H_{w,2}$ in $E(\F_q)$ containing $p_0$. These middle codes are locally recoverable with availability $2$, the two recovery sets being the non-trivial cosets of $G_{v,1}$ and $G_{v,2}$ in $E(\F_q) $ containing $p_0$. The code has the following parameters.
    \begin{itemize}
        \item For the lower codes:
        \[n_{2,1} = n_{2,2}=v, k_{2,1}=k_{2,2}=v-1, d_{2,1}=d_{2,2}=2.\]
        \item For the middle codes:
        $$n_{1,1}=n_{1,2} = v^2 \cdot w,$$ $$k_{1,1}=k_{1,2} = (v-1)^2 \cdot (w-1),$$
        $$d_{1,1} \geq n_{1,1}  - Z_{p_0,1,1}  = 8,
        $$
        $$d_{1,2} \geq n_{1,2}- Z_{p_0,1,2}= 8,
        $$ 
        where $Z_{p_0,1,i}$ is the maximum number of zeros of a function contained in $V$ on the evaluation set of the middle code $\C_{1,i}$ which includes the point $p_0$ in its support.
        \item For the full code: $$n= \ell \cdot m^2,$$ $$k=t \cdot (v-1)^2 \cdot (w-1)^2,$$ 
        $$d \geq n - Z_{\text{global}}  = 16(\ell - t).$$
        Here, $Z_\text{global}$ is the maximum number of zeros of a function in $V$ on the evaluation set $T$ of the global code. 
    \end{itemize}
\end{prop}

\begin{myproof}
    The lengths and the dimensions of the lower, middle and full codes can be easily obtained from the construction. In fact, $\C(E,V,T)$ satisfies \Cref{def: H-level hierarchy with availability}, with recovery procedure given by interpolation: indeed, once again, it can be traced back to a univariate or multivariate interpolation problem, since the restrictions of $V$ to the evaluation sets of the middle and lower codes become respectively
    $$V_1 = \langle f_1^{a_1} \cdot f_2^{a_2} \cdot g_2^{b_2} \rangle, \ \ \ \ \ \ V_2 = \langle f_1^{a_1} \cdot f_2^{a_2} \cdot g_1^{b_1} \rangle$$
    and 
    $$V'_1 = \langle f_2^{a_2} \rangle, \ \ \ \ \ \ V'_2 = \langle f_1^{a_1} \rangle.$$
    The conditions on $f_i$ and $g_i$ are crucial for a correct application of the interpolation procedure.
    To give some estimates on the minimum distances $d_{1,i}$ and $d$, the proof consists of estimating the maximum number of zeros of a function $\mathcal{F}$ in $V_i$ and in $V$. We work only on the middle codes and the vector space $V_1$, counting the maximum number of zeros layer by layer through the hierarchical structure: in fact, the evaluation set of the middle code $\C_{p_0,1,1}$ supported on a point $p_0 \in T$ is its coset in $E(\F_q)$ with respect to $H_{w,1}$, which can be seen as a disjoint union of $w$ distinct cosets of $E[v]$. As a consequence of the condition \ref{cg23}, $g_2$ assumes exactly $w$ different values on these cosets. Moreover, a function $\mathcal{F} \in V_1$ corresponds to a polynomial in $g_2$ of maximum degree $w-2$, thus, it can vanish on at most $w-2$ of these cosets, contributing to at most $v^2 \cdot (w-2)$ zeros. On the remaining two $E[v]$-cosets, $\mathcal{F}$ is clearly not identically zero and, on each one of these blocks, it can be seen as a polynomial in $f_1$ and $f_2$. Every $E[v]$-coset is the disjoint union of $v$ distinct $G_{v,1}$-cosets. Condition \ref{cf23} implies that the function $f_2$ assumes exactly $v$ distinct values across them and, again, since $\mathcal{F}$ is a polynomial in $f_2$ of maximum degree $v-2$, it can vanish on at most $v-2$ $G_{v,1}$-cosets for each $E[v]$-cosets, contributing to $2v \cdot (v-2)$ zeros. We can repeat this argument for the $G_{v,2}$-cosets and the function $f_1$, obtaining other $4 \cdot (v-2)$ zeros. Summing these contributions, we have the following estimates for the minimum distance of $\C_{p_0,1,1}$:
    $$d_{1,1} \geq n_{1,1} - Z_{p_0,1,1} =  v^2 \cdot w - v^2 \cdot (w-2) - 2v \cdot (v-2) - 4 (v-2) = 8.$$
    The proof is the same for the middle code $\C_{p_0,1,2}$, obtaining the same result. For the global minimum distance, we can use the same arguments, including the contribution of the functions $h_i$: this leads to the following estimate:
    \begin{align*}
    d &\geq n - Z_{\text{global}}  \\
    &=  \ell \cdot v^2 \cdot w^2 - t \cdot v^2 \cdot w^2 - (\ell - t) \cdot \big[v^2 \cdot w \cdot (w-2) - 2v^2 \cdot (w-2) - 4v \cdot (v-2) - 8 \cdot (v-2)\big] \\
    &= 16 \cdot (\ell - t).
    \end{align*}
   This concludes the proof.
\end{myproof}

\begin{rem}
    The hierarchical construction presented in this section can be interpreted as a particular case of product codes. Indeed, the evaluation set $T$ is structured as a Cartesian product of nested cosets and the vector space $V$ is spanned by the product of algebraically independent functions $\{h_i, f_1, f_2, g_1, g_2\}$. 
    As a direct consequence, the global minimum distance $d$ matches the product of the minimum distances of the constituent local codes. This structural property is a well-known feature in algebraic coding theory, which is presented and discussed in \cite[Ch.~18]{MS77}.
\end{rem}

\subsection{Hierarchical locality and availability on elliptic surfaces}\label{construction_surfaces: section 5}

This subsection is the analogous version of Subsection \ref{construction_surfaces: section 4}, to obtain hierarchical locality and availability, working on the generic fiber of an elliptic surface. We will not repeat the entire construction: our strategy is to adopt the same ideas developed in \Cref{construction_surfaces: section 4}, applying them to the setup introduced in \Cref{construction_curves: section 5}

Consider a projective curve $B$ of genus $g_B$ defined over a finite field $\F_q$ and an elliptic $\F_q$-surface $\pi: \E \longrightarrow B$, with generic fiber $E/K$, where $K=\F_q(B)$. Given an integer $m=v \cdot w$ coprime with $q$, we assume that $E[m] \subseteq E(K)$ and we define the groups $G_{v,i}$, $G_{w,i}$, the rational functions $f_i, g_i, h_j \in K(E)$ and the evaluation set $\tilde{T} \subseteq E(K)$ as in \Cref{construction_curves: section 5}.

To define the evaluation set on the surface $\E$, we proceed analogously to \Cref{construction_surfaces: section 4}: we choose a subset $C \subseteq B(\mathbb{F}_q)$ of good points by avoiding the roots of the discriminant $\Delta$ of the generic elliptic fiber, the poles of the coefficients defining $f_i, g_i$, and $h_j$, and the roots of the separating functions $\delta_{f_i, p, s}$ and $\delta_{g_i, p, s}$. The existence of an appropriate set $C$ is guaranteed as discussed in \Cref{rem:bad_points}.

The global evaluation set on the elliptic surface is defined as the disjoint union of the specialized local evaluation sets:
\begin{equation*}
    T = \bigcup_{\gamma \in C} T_\gamma
\end{equation*}
where $ T_\gamma = \{ sp_\gamma(p) \in E_\gamma(\mathbb{F}_q) \mid p \in \tilde{T}$, and $sp_\gamma : E(K) \longrightarrow E_\gamma(\mathbb{F}_q)$ is the specialization map.

\begin{thm}\label{prop:final_params}
Let $D \in Div_{\F_q}(B)$ be an effective divisor of degree $\delta$, with $g_B \le \delta < \#C$, with support disjoint from the set $C$. Let $V \subseteq \mathbb{F}_q(\mathcal{E})$ be the $\mathbb{F}_q$-vector space defined by:
\begin{equation*}
    V = \mathrm{Span} \left\{ b \cdot h_i \cdot f_1^{a_1} f_2^{a_2} g_1^{b_1} g_2^{b_2} \mid b \in H^0(B, \mathcal{O}_B(D)), \, 1 \le i \le t, \, 0 \le a_j \le v-2, \, 0 \le b_j \le w-2 \right\}.
\end{equation*}
The associated AG code $\mathcal{C}(\mathcal{E}, V, T)$ inherits the 2-level hierarchy with availability $(2, 2)$ on every elliptic fiber $E_\gamma$ for $\gamma \in C$, with the same construction presented in \Cref{prop: parameters_curve section 5}. Its global parameters satisfy:
\begin{itemize}
    \item $n = \#C \cdot \ell \cdot v^2 w^2$,
    \item $k = h^0(B, D) \cdot t \cdot (v-1)^2 \cdot (w-1)^2$,
    \item $d \ge (\#C - \delta) \cdot 16(\ell - t)$.
\end{itemize}
\end{thm}

\begin{myproof}
By construction, for any fixed $\gamma \in C$, the evaluation map restricted to the fiber $E_\gamma$ is well-defined, and the punctured code $\mathcal{C}(E_\gamma, V_\gamma, T_\gamma)$ coincides with the code constructed in the previous subsection, providing the required local hierarchy and availability. The global length $n$ and the dimension $k$ follow directly from the structure of $T$ and $V$, respectively. Finally, the global minimum distance is bounded from below by taking the minimum distance on a single fiber, which is $16(\ell-t)$, as computed in \Cref{prop: parameters_curve section 5}, multiplied by the minimum number of active fibers $(\#C - \delta)$, following the same counting argument presented in the proof of \Cref{prop: parameters_surfaces section 4}.
\end{myproof}

\section{Conclusion}\label{sec: conclusion}
In this paper, we presented some new constructions of LRCs from elliptic surfaces, which allow us to get availability, hierarchical locality and the combination of the two. Our first construction is a generalization and systematization of \cite[\S VI]{SVV}, leading to the achievement of locality with availability strictly greater than $2$ from elliptic surfaces, defined over an arbitrary base curve. The fibered structure is used to build the evaluation set and the vector space, but it does not give the code hierarchical locality, since the restriction of the code to every elliptic curve of the form $E_\gamma$, with $\gamma$ contained in the base curve, is not effectively an LRC. 

The constructions developed in Sections \ref{section: HLRC elliptic surfaces} and \ref{section: HLRC with availability} are intrinsically distinct from the one detailed in Section \ref{section: availability ell surfaces}. Rather than working directly with the geometry of the surface, these constructions originate from the study of elliptic curves and their sequences of quotients, and are lifted to elliptic surfaces via the generic fiber. 

Another difference lies in how the recovery capabilities of the codes are established and proven to work. In Section \ref{section: availability ell surfaces}, the evaluation set and the spaces of rational functions are chosen to satisfy the hypotheses of Abel's Theorem, which guarantees the existence of several disjoint recovery sets. In Sections \ref{section: HLRC elliptic surfaces} and \ref{section: HLRC with availability}, the repair procedure, which relies on polynomial interpolation, is made possible by the algebraic conditions imposed on the evaluations of the rational functions: mainly the requirements outlined in Conditions \ref{1c)} and \ref{2c)}. We ensure that the interpolation remains well-posed, while avoiding restrictive limitations on the evaluation set itself.

Although these constructions are made on elliptic surfaces, they are strongly based on the rich spectrum of arithmetic properties of elliptic curves. The geometry of the surface is used to introduce a multi-dimensional setting, with two or more levels for the recovery patterns. 

Finally, the estimates of the parameters have been carried out using techniques already explored in the literature on curves and fibered surfaces. Better estimates can be performed, for example, by taking into consideration specific equations for the definition of the algebraic varieties.

\subsection{Further developments}
We conclude by offering some ideas for further developments, starting from the results and the constructions presented in this paper.

Investigating specific algebraic conditions to impose on the evaluation functions could be a first step forward: by working with the underlying Riemann--Roch spaces, one could force their zeros to fall outside the evaluation set. This approach may lead to increasing the minimum distance of the resulting codes. 

The focus of this paper is on elliptic surfaces. Local properties of codes from ruled surfaces and other fibered surfaces have been recently studied in \cite{BlaHal25,SalVic25}. It would be interesting to investigate other fibered surfaces, of general type, for instance, to construct LRCs with additional properties and/or improved parameters. 

Finally, the other tool we have extensively used in our work is the group structure of elliptic curves.  It is natural to ask whether our work can be generalized to abelian varieties of higher dimension. For instance, codes from abelian surfaces have already been studied in \cite{ABHP21}, but their local properties have not been investigated yet. Abelian varieties could serve as a good source of LRCs with additional properties, considering their rich group structure on the set of points. 

\nocite{}
\bibliographystyle{alpha}\bibliography{mybib}

@article{ABHP21,
  title={Algebraic geometry codes over abelian surfaces containing no absolutely irreducible curves of low genus},
  author={Aubry, Yves and Berardini, Elena and Herbaut, Fabien and Perret, Marc},
  journal={Finite Fields and Their Applications},
  volume={70},
  pages={101791},
  year={2021},
  publisher={Elsevier}
}

@article{ABHP21b,
  title={Bounds on the minimum distance of algebraic geometry},
  author={Aubry, Yves and Berardini, Elena and Herbaut, Fabien and Perret, Marc},
  journal={Arithmetic, geometry, cryptography and coding theory. Contemp. Math},
  volume={770},
  pages={11},
  year={2021},
  publisher={American Mathematical Soc.}
}

@article{ballentine2019,
  title={Codes with hierarchical locality from covering maps of curves},
  author={Ballentine, Sean and Barg, Alexander and Vl{\u{a}}du{\c{t}}, Serge},
  journal={IEEE Transactions on Information Theory},
  volume={65},
  number={10},
  pages={6056-6071},
  year={2019},
  publisher={IEEE}
}

@incollection{BargEtAl2017,
  title={Locally recoverable codes from algebraic curves and surfaces},
  author={Barg, Alexander and Haymaker, Kathryn and Howe, Everett W and Matthews, Gretchen L and V{\'a}rilly-Alvarado, Anthony},
  booktitle={Algebraic Geometry for Coding Theory and Cryptography: IPAM, Los Angeles, CA, February 2016},
  pages={95-127},
  year={2017},
  publisher={Springer}
}

@article{BargTamoEtAl2017,
  title={Locally recoverable codes on algebraic curves},
  author={Barg, Alexander and Tamo, Itzhak and Vl{\u{a}}du{\c{t}}, Serge},
  journal={IEEE Transactions on Information Theory},
  volume={63},
  number={8},
  pages={4928-4939},
  year={2017},
  publisher={IEEE}
}

@article{BlaHal25,
  title={Ruled surfaces over finite fields, and some codes over them},
  author={Blache, R{\'e}gis and Hallouin, Emmanuel},
  journal={arXiv preprint arXiv:2509.18698},
  year={2025}
}

@article{BergEtAl2024,
  title={{Codes with hierarchical locality on Artin--Schreier surfaces}},
  author={Berg, Jennifer and Malmskog, Beth and West, Mckenzie},
  journal={Journal of Algebra and Its Applications},
  volume={24},
  number={13n14},
  pages={2541023},
  year={2025},
  publisher={World Scientific}
}

@article{CouvreurEtAl2020,
  title={Toward good families of codes from towers of surfaces},
  author={Couvreur, Alain and Lebacque, Philippe and Perret, Marc},
  journal={Arithmetic, geometry, cryptography and coding theory. Contemp. Math},
  volume={770},
  pages={59-93},
  year={2021}
}

@article{goppa1983,
  title={Algebraico-geometric codes},
  author={Goppa, Valerii Denisovich},
  journal={Mathematics of the USSR-Izvestiya},
  volume={21},
  number={1},
  pages={75},
  year={1983},
  publisher={IOP Publishing}
}

@book{gortz2020,
  title={Algebraic geometry I: schemes},
  author={G{\"o}rtz, Ulrich and Wedhorn, Torsten},
  year={2020},
  publisher={Springer}
}

@book{Hartshorne,
  title={Algebraic geometry},
  author={Hartshorne, Robin},
  volume={52},
  year={2013},
  publisher={Springer Science \& Business Media}
}

@article{Malmskog2018,
  title={Locally recoverable codes with availability  $t\geq2$ from fiber products of curves},
  author={Haymaker, Kathryn and Malmskog, Beth and Matthews, Gretchen L},
  journal={Advances in Mathematics of Communications},
  volume={12},
  number={2},
  pages={317--336},
  year={2018}
}

@article{Malmskog2025,
  title={Algebraic hierarchical locally recoverable codes with nested affine subspace recovery},
  author={Haymaker, Kathryn and Malmskog, Beth and Matthews, Gretchen},
  journal={Designs, Codes and Cryptography},
  volume={93},
  number={1},
  pages={111-132},
  year={2025},
  publisher={Springer}
}

@article{rawat2016,
  title={Locality and availability in distributed storage},
  author={Rawat, Ankit Singh and Papailiopoulos, Dimitris S and Dimakis, Alexandros G and Vishwanath, Sriram},
  journal={IEEE Transactions on Information Theory},
  volume={62},
  number={8},
  pages={4481--4493},
  year={2016},
  publisher={IEEE}
}

@inproceedings{sasidharan2015,
  title={Codes with hierarchical locality},
  author={Sasidharan, Birenjith and Agarwal, Gaurav Kumar and Kumar, P Vijay},
  booktitle={2015 IEEE International Symposium on Information Theory (ISIT)},
  pages={1257--1261},
  year={2015},
  organization={IEEE}
}

@book{silverman2009,
  title={The arithmetic of elliptic curves},
  author={Silverman, Joseph H},
  volume={106},
  year={2009},
  publisher={Springer}
}

@book{silverman2013advanced,
  title={Advanced topics in the arithmetic of elliptic curves},
  author={Silverman, Joseph H},
  volume={151},
  year={2013},
  publisher={Springer Science \& Business Media}
}

@article{SVV,
  title={Locally recoverable codes on surfaces},
  author={Salgado, Cec{\'\i}lia and V{\'a}rilly-Alvarado, Anthony and Voloch, Jos{\'e} Felipe},
  journal={IEEE Transactions on Information Theory},
  volume={67},
  number={9},
  pages={5765--5777},
  year={2021},
  publisher={IEEE}
}

@article{TamoBarg2014,
  title={A family of optimal locally recoverable codes},
  author={Tamo, Itzhak and Barg, Alexander},
  journal={IEEE Transactions on Information Theory},
  volume={60},
  number={8},
  pages={4661-4676},
  year={2014},
  publisher={IEEE}
}

@article{tamo2016bounds,
  title={Bounds on the parameters of locally recoverable codes},
  author={Tamo, Itzhak and Barg, Alexander and Frolov, Alexey},
  journal={IEEE Transactions on information theory},
  volume={62},
  number={6},
  pages={3070--3083},
  year={2016},
  publisher={IEEE}
}

@book{AGBook,
  title={Algebraic geometric codes: basic notions},
  author={Vladut, Serge and Nogin, Dmitry and Tsfasman, Michael},
  year={2007},
  publisher={American Mathematical Society}
}

@book{MS77,
  title={The Theory of Error-Correcting Codes},
  author={MacWilliams, Florence Jessie and Sloane, Neil James Alexander},
  volume={16},
  year={1977},
  publisher={North-Holland}
}

@article{SalVic25,
  title={Locally Recoverable Codes with availability from a family of fibered surfaces},
  author={Salgado, Cec{\'\i}lia and Vicino, Lara},
  journal={arXiv preprint arXiv:2512.08100},
  year={2025}
}

@article{kamath2013codeslocalregeneration,
  title={Codes with local regeneration and erasure correction},
  author={Kamath, Govinda M and Prakash, N and Lalitha, V and Kumar, P Vijay},
  journal={IEEE Transactions on information theory},
  volume={60},
  number={8},
  pages={4637--4660},
  year={2014},
  publisher={IEEE}
}

@book{lang2012algebra,
  title={Algebra},
  author={Lang, Serge},
  year={2012},
  publisher={Springer Science \& Business Media}
}
\end{document}